\PassOptionsToPackage{table,svgnames}{xcolor}

\documentclass[cameraready]{vgtc}

\graphicspath{{figures/}{pictures/}{images/}{./}} 

\usepackage{times}                     

\usepackage{tabu}                      
\usepackage{booktabs}                  
\usepackage{lipsum}                    
\usepackage{mwe}                       
\usepackage{makecell}

\usepackage{mathptmx}                  

\usepackage{subfigure}
\usepackage{subcaption}

\newcommand{\etal}{et~al.}

\newcommand{\xref}[1]{\cref{#1}}

\newcommand{\new}[1]{{\color{black}#1}}

\def\code#1{\texttt{#1}} 

\usepackage{amsmath}
\usepackage{float}
\usepackage{color}
\usepackage{booktabs}

\usepackage{array}
\newcolumntype{P}[1]{>{\centering\arraybackslash}m{#1}}

\onlineid{1026}

\vgtccategory{Research}

\vgtcinsertpkg

\title{Lagrangian Simulation Volume-Based Contour Tree Simplification}

\author{Domantas Dilys\thanks{e-mail: D.Dilys@leeds.ac.uk}\\ %
      \parbox{1.4in}{\scriptsize \centering University of Leeds \\ School of Computer Science} %
\and Hamish Carr\thanks{e-mail: H.Carr@leeds.ac.uk}\\ %
     \parbox{1.4in}{\scriptsize \centering University of Leeds \\ School of Computer Science} %
\and Steven Boeing\thanks{e-mail: S.Boeing@leeds.ac.uk}\\ %
     \parbox{1.4in}{\scriptsize \centering University of Leeds \\ School of Earth and Environment}}

\teaser{
\centering
    \subfigure[$1024^3$ grid volume rendering]{\includegraphics[width=0.24\textwidth]{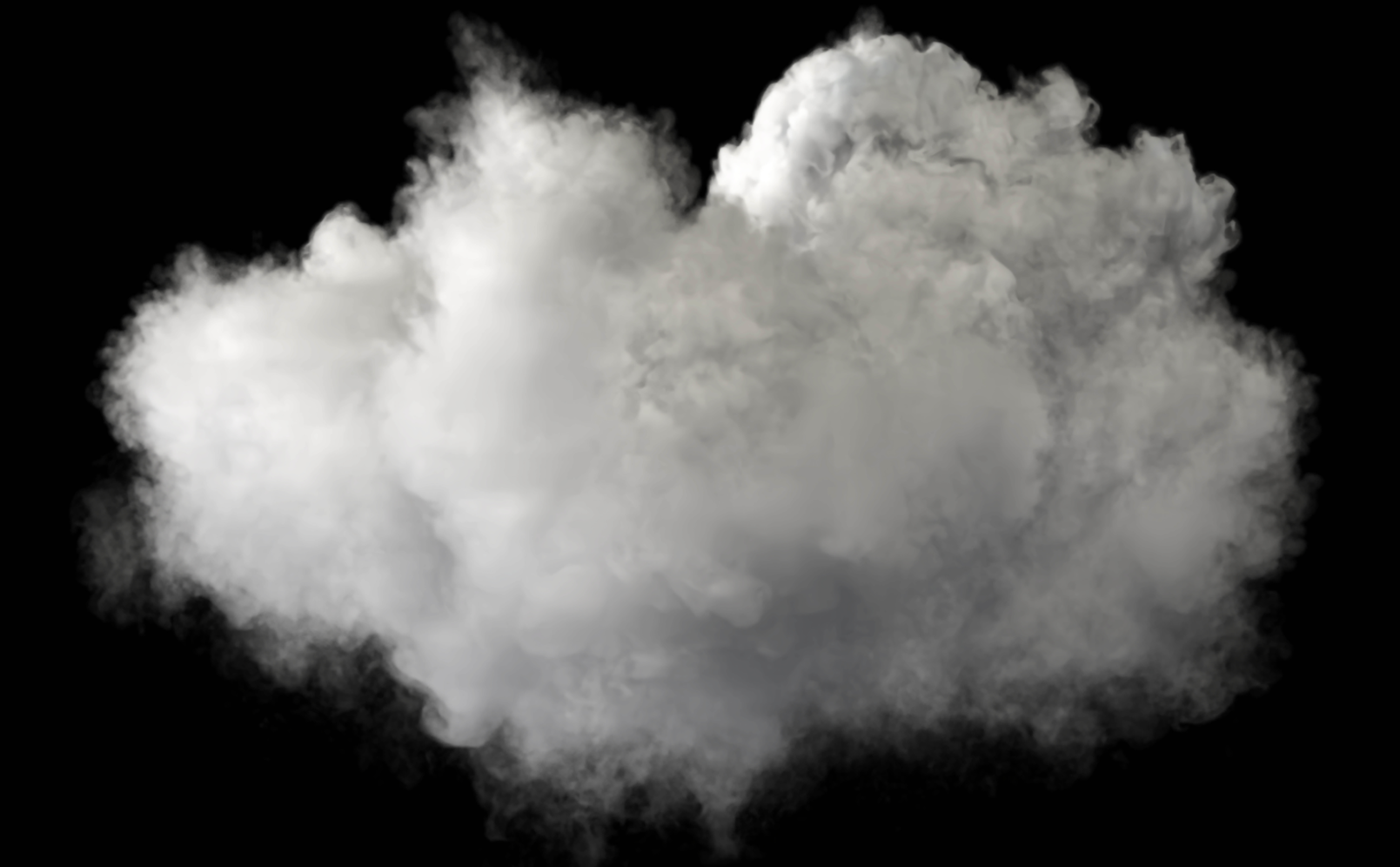}} 
    \subfigure[$512^3$ grid segmentation]{\includegraphics[width=0.24\textwidth]{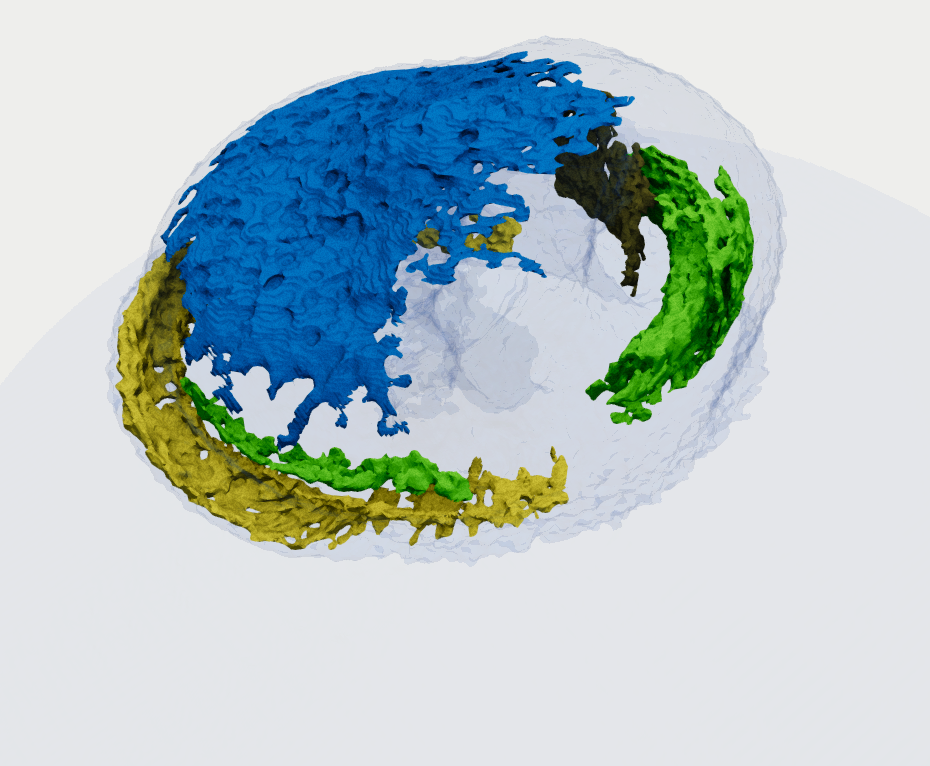}} 
    \subfigure[2M parcel Delaunay mesh ]{\includegraphics[width=0.24\textwidth]{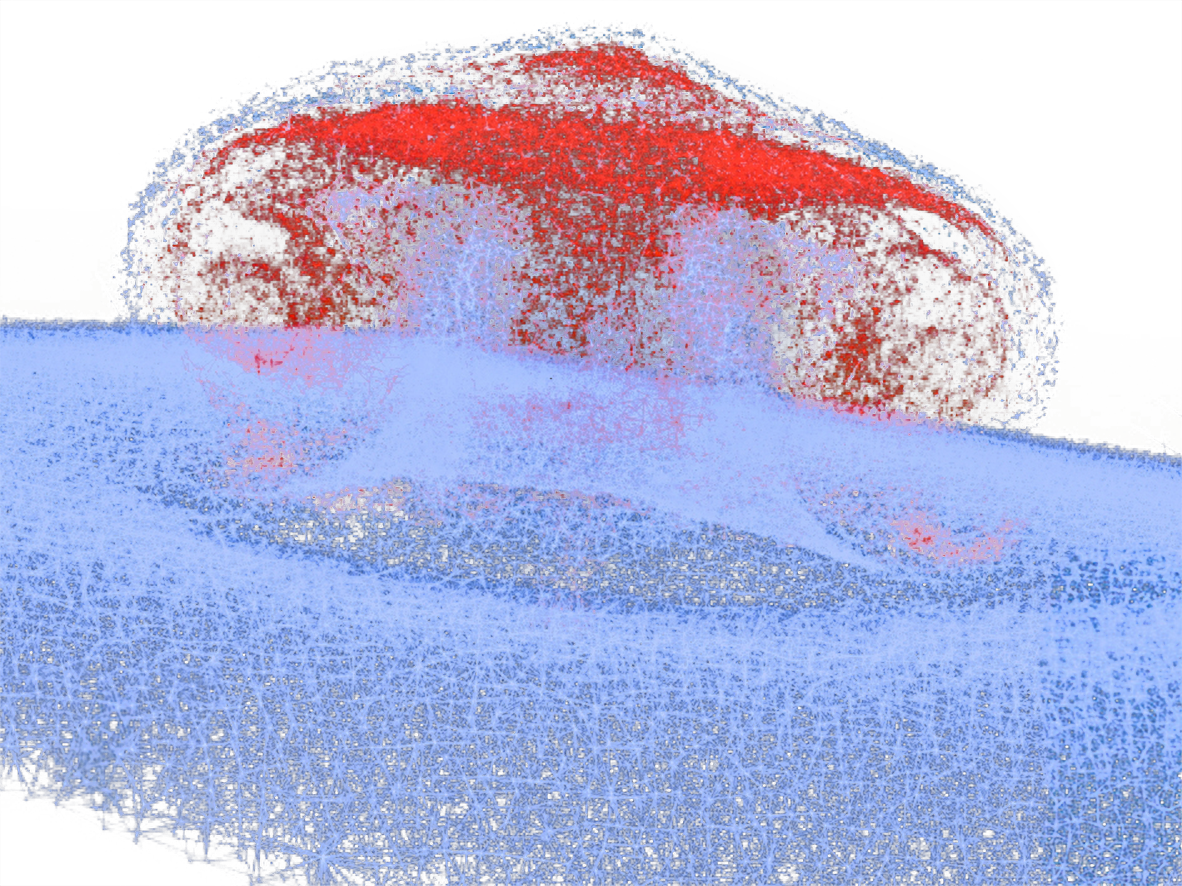}}
    \subfigure[2M parcel segmentation ]{\includegraphics[width=0.24\textwidth]{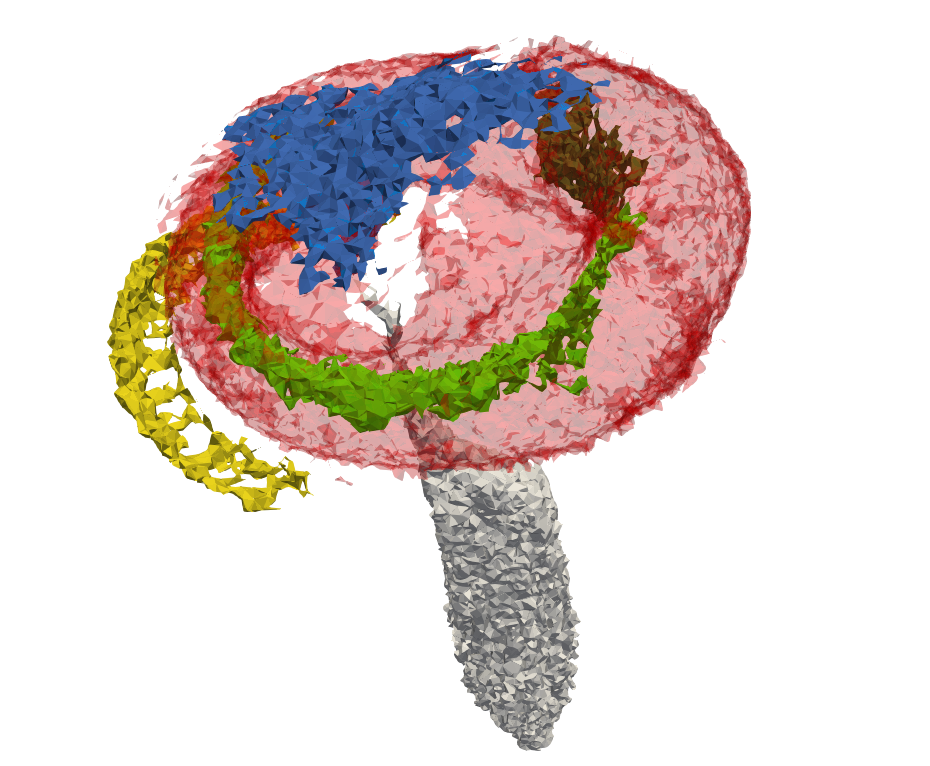}}
    \caption{Visualisation and topological analysis of Lagrangian simulations often have to rely on costly resampling onto a grid (a), (b). We perform topological data analysis directly from the parcel neighbourhoods (c) faster, and with good quality (d). }
  \label{fig:teaser}
}

\abstract{
Many scientific and engineering problems are modelled by simulating scalar fields defined either on space-filling meshes (\textit{Eulerian}) or as particles (\textit{Lagrangian}). 
For analysis and visualization, topological primitives such as contour trees can be used, but these often need simplification to filter out small-scale features. 
For parcel-based convective cloud simulations, simplification of the contour tree requires a volumetric measure rather than persistence. 
Unlike for cubic meshes, volume cannot be approximated by counting regular vertices. 
Typically, this is addressed by resampling irregular data onto a uniform grid. 
Unfortunately, the spatial proximity of parcels requires a high sampling frequency, resulting in a massive increase in data size for processing.
We therefore extend volume-based contour tree simplification to parcel-in-cell simulations with a graph adaptor in \code{Viskores} (\code{VTK-m}), using Delaunay tetrahedralization of the parcel centroids as input. 
Instead of relying on a volume approximation by counting regular vertices – as was done for cubic meshes – we adapt the 2D area splines reported by Bajaj \etal~\cite{BPS97}, and Zhou~\etal ~\cite{zhou2018efficient}.
We implement this in \code{Viskores} (formerly called \code{VTK-m}) as prefix-sum style hypersweeps for parallel efficiency and show how it can be generalized to compute any integrable property.
Finally, our results reveal that contour trees computed directly on the parcels are orders of magnitude faster than computing them on a resampled grid, while also arguably offering better quality segmentation, avoiding interpolation artifacts.
} 


\begin{document}

\firstsection{Introduction}

\maketitle

Scientific computing and engineering depend on fluid simulations, which are increasingly computed using \textit{Lagrangian }techniques for efficiency instead of \textit{Eulerian} rectilinear grids.  
Once the data are computed, it must still be visualized, understood, and analyzed.  

Contour trees help to visualize scientific datasets and can be computed directly on irregular tetrahedral meshes. 
However, implementations have focused on regular rectilinear grids. 
For topological simplification, the standard practice to compute the volume has been by simple vertex count approximation, which is a cheap operation and converges rapidly as the grid resolution increases~\cite{carr2004simplifying}.

\textit{Parcel-in-Cell} (PIC) are hybrid Lagrangian-Eulerian simulations, tracking fluid \emph{parcels} of nonzero volume (unlike \textit{particles}), driven by an underlying fixed-resolution coarse \textit{simulation grid}. 
Visualization of PIC data needs an additional post-processing step, computing the weighted contributions of every parcel at each point on the \textit{interpolation grid} (much finer than the \textit{simulation grid}).
This increases costs in storage, computation, visualization, and analysis.  

Because parcels are often tightly packed in space, the \textit{Nyquist limit}~\cite{shannon1949communication} requires a high sampling rate, and working directly with the parcels is preferable. 
Since PIC simulations exchange material between adjacent parcels, we construct the Delaunay tetrahedralization of the parcel centroids and use this as an irregular mesh.

The approximation of volume by vertex count is only appropriate for regular meshes~\cite{carr2004topological}. 
For irregular meshes, we revert to the observation by Pascucci~\etal~\cite{BPS97} that geometric properties, such as volume, can be expressed by sets of polynomial coefficients.

Initial \code{VTK-m} (now \code{Viskores})~\cite{MSU16} contour tree implementation made the strong assumption that vertex degree was bounded (as is normal in regular grids) and applied optimizations on this basis. 
We report on the changes necessary to compute contour trees for any irregular mesh, complete with hypersweep-based summation of polynomial coefficients, branch decomposition, and visualizations.

We show that working directly from the parcels is more efficient than resampling at high resolutions, and also more accurate.

\begin{figure*}
\centerline{\includegraphics[width=.9\linewidth]{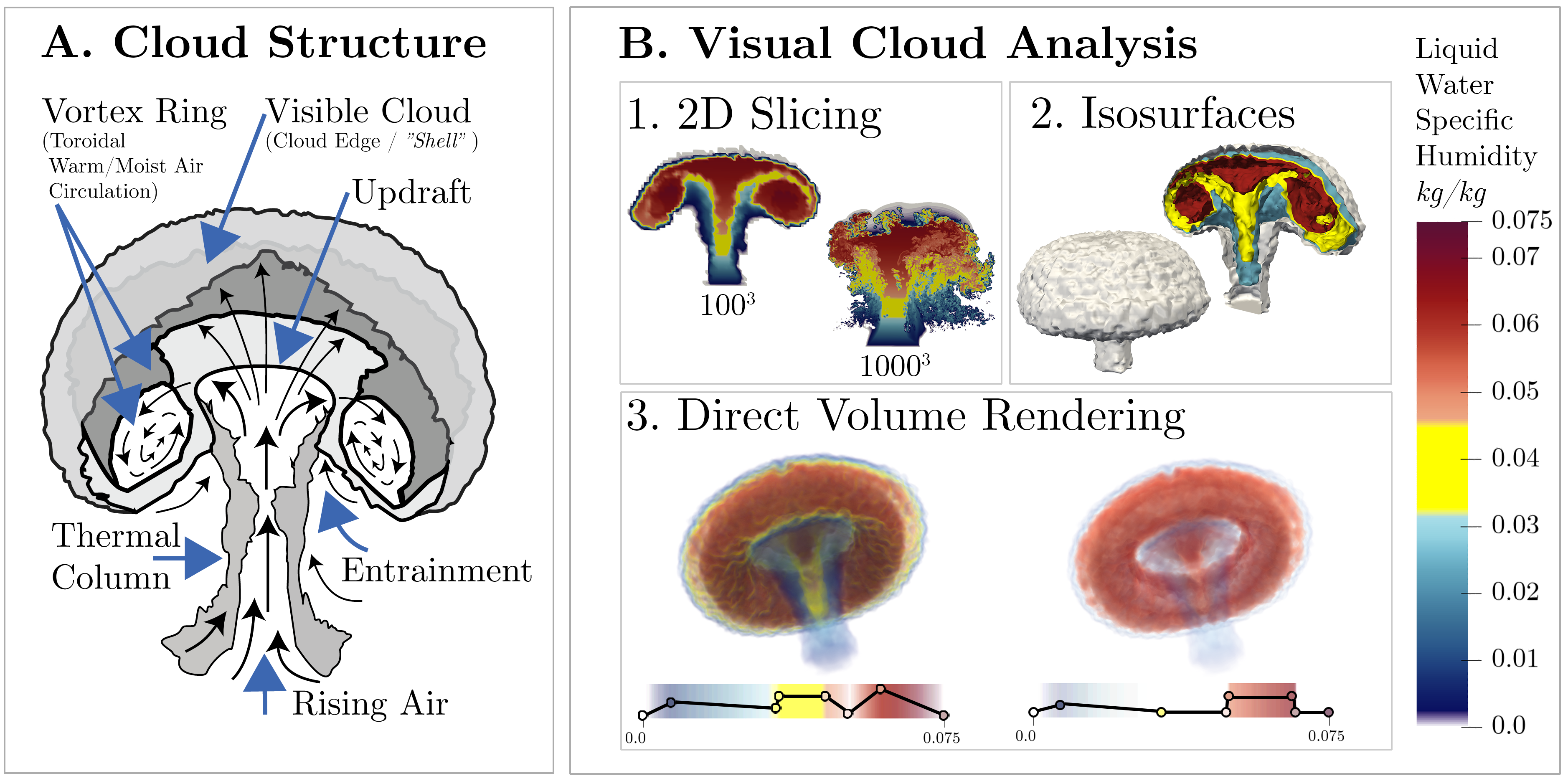}}
\caption{Current visual cloud structure analysis techniques. \textit{ 2D slicing} (B.1) is a common way for inspecting vortex rings (A), but features tend to spread out at higher resolutions (e.g. at $1000^3$ grids). \textit{Isosurfaces} (B.2) often hide structures of interest with their outer shell (left), while cross-sections (right) show the characteristic $\Phi$ profile but miss the full 3D form. \textit{Direct Volume Rendering} (B.3) is somewhat better, but the blurry nature of  rendering makes it hard to perceive features of interest (left), and transfer function design is complex (bottom-right).}
\label{fig:cloud-structure-analysis}
\end{figure*}

We start with a brief description of the problem in meteorology \xref{sec:meteorology}, and a summary of the differences between grid-based and Parcel-in-Cell simulations in \xref{sec:parcelincell}, followed by a summary of the key developments in computational topology relevant to this paper in \xref{sec:contourtreeanalysis}.  We will then be able to state the problem being solved in this paper in \xref{sec:problemstatement}.  We will then describe the details of the polynomial coefficients necessary to compute volume in a tetrahedral mesh in \xref{sec:polynomialcoefficients}, and how to adapt the parallel hypersweep to them in \xref{sec:polynomialhypersweep}.  Once this is complete, we will discuss the implementation details for this pipeline in the \code{Viskores} toolkit in \xref{sec:implementation}, and evaluate the performance and quality in \xref{sec:results}. Finally, we will summarize our results and speculate on future work in \xref{sec:summary}. 


\section{Meteorological Background}
\label{sec:meteorology}

Cloud formation depends on transport of air from the surface to upper levels. 
This \emph{thermal column} of warm air can curl up into \emph{vortex rings} as it punches into colder surrounding air (\xref{fig:cloud-structure-analysis} A).

\textit{Entrainment}, a process of dry surrounding air entering a moist cloud, influences cloud droplet distribution~\cite{blyth1980influence} and cloud lifespan~\cite{mellado2017cloud}. 
This mixing process can speed up when vortex rings are present~\cite{dabiri2004fluid}. 
If the thermal cools enough during the intake of drier air, the cloud can dissipate. 
Therefore, the cloud life-cycle depends on the entrainment rate, which itself is an active area of research~\cite{blyth1993entrainment}.

Vortex rings and related mixing processes are studied to understand entrainment better.
High-resolution data are essential for capturing local variation and detailed flow structure~\cite{frey2022epic}.
Vortex rings are often identified \emph{qualitatively}, through visualization (\xref{fig:cloud-structure-analysis} B). 
In 2D cross-sections, they appear as \textbf{$\Phi$}-shaped structures, with the thermal column in the middle and circles on each side. 
In 3D, they form tori that trap air with distinct scalar values often unknown beforehand. 
Both views are unreliable, as the ring’s appearance varies with transfer functions, simulation parameters, and resolution~\cite{boing2019comparison}.

Although these visualizations are effective at small scale for deriving insight, the long-term goal is reliable \emph{quantitative} detection over large regions, such as west Africa. 
Moreover, spatial metrics, such as the total volume of the vortex ring, are useful for modeling. 
One of the possible solutions for a more robust analytic pipeline is to apply contour tree analysis (\xref{sec:contourtreeanalysis}). 
Before we approach this, however, we must look at one of the key characteristics of modern meteorological simulations: the use of parcel-in-cell simulations.

\section{Parcel-in-Cell Simulations}
\label{sec:parcelincell}

Historically, meteorological simulation started with Eulerian approaches on regular grids in 2D and 3D, as with many scientific and engineering problems~\cite{deardorff1970three}.
Regular grids have uniform resolution, even in inactive regions.
While effective, ever-finer scales require increasing resolution, driving the computational cost up for \textit{Large-Eddy Simulation} (LES) and \textit{direct numerical simulation}. 

Regular grids are not the only way.
Lagrangian methods track individual flow elements, focusing on active regions. Early examples include \textit{contour advection} – following the fluid boundaries (like the outline of an ink drop).
Later, to combine the strengths of both, hybrid methods were derived, such as the \textit{Parcel-in-Cell} (PIC), where volumetric \emph{parcels} are tracked through an underlying \textit{simulation grid}. The parcels carry prognostic information on dynamic (vorticity) and thermodynamic (heat, moisture) properties of the flow at a scale that is finer than the simulation grid. The \textit{Elliptical Parcel-in-Cell} (EPIC) variant improves spatial packing (incompressibility) and the accuracy of advection by using deformable ellipses instead of spheres~\cite{frey2022epic}. Earlier studies~\cite{frey2022epic, frey20233d} showed that EPIC compares favorably with gridded LES models run at even slightly higher resolutions, with EPIC producing much less diffuse vortex rings than LES~\cite{frey20233d} because diffusion can be explicitly controlled in EPIC.



Advances in PIC and EPIC have not always been accompanied by advances in the visualization tools for examining and analyzing the data. As a result, visualization is currently performed largely by resampling the simulation to a high-resolution grid, then invoking existing tools: as we will see in \xref{sec:results}, this is not the ideal solution.

\section{Contour Tree Analysis}
\label{sec:contourtreeanalysis}
Although both isosurface extraction~\cite{lorensen1987marching} and direct volume rendering \cite{levoy1988display,Drebin1988} assist in analysing cloud formation, selecting suitable isosurfaces is  difficult, as the features of interest are inside the cloud and are likely to be occluded by exterior structures (\xref{fig:cloud-structure-analysis} B.2).

One solution is the \emph{flexible isosurface}~\cite{CSV10}, which uses the \emph{contour tree} to select connected components of isosurfaces (\emph{contours}), allowing finer-grained choice of what is seen. 
Moreover, geometric properties of contours such as volume, surface area, or integrals of secondary properties can be pre-computed with the tree.

A Reeb graph~\cite{Ree46} is the quotient space under continuous contraction of connected components of inverse images of an isovalue $h$ in a function $f$ on some manifold $M$.  
Most often, the manifold $M$ is a mesh from a simulation, and is a compact subset of either $\mathcal{R}^2$ or $\mathcal{R}^3$. 
In this case there are no cycles, and the Reeb graph is referred to as a \emph{contour tree}, composed of \emph{supernodes} and \emph{superarcs}, with the mesh vertices as \emph{regular nodes} along the superarcs.

The contour tree is useful for extracting isosurfaces~\cite{BPS97} or individual contours, but also for finding features and their relationships based on the extrema and critical points~\cite{CSV10}. 
They can be computed efficiently in serial for simplicial meshes~\cite{CSV10}, non-simplicial meshes~\cite{PC03,CS09} or by choosing a suitable \emph{topology graph} that captures neighborhood relationships~\cite{CS09,correa2011towards}. 
More recently parallel algorithms have been described for shared memory~\cite{GFJ16}, PRAM~\cite{CWS16}, distributed~\cite{PC03} or hybrid distributed/PRAM~\cite{CRW22} models.

As data sizes increase, so do contour trees, which can be simplified by discarding superarcs with small height~\cite{PCS04}, or by using geometric properties such as volume in a similar fashion~\cite{CSV10}.

While edge height is easy to calculate, other geometric properties are not.
Bajaj~\etal~\cite{BPS97} showed that geometric properties such as volume, surface, etc. are functions of the isovalue, and that in simplicial meshes, are often per-simplex polynomial splines.
Then, one can compute the function for every possible isovalue in advance. 
However, their development of the 3D are splines was slightly imprecise, and did not apply Federer's co-area formula~\cite{federer_gmt_69} to adjust for gradient which was identified as necessary~\cite{SSD08}, with Zhou \etal \cite{zhou2018efficient} confirming this and correcting the missing details.

Later work~\cite{CSV10} established that these computations could be extended from functions for the entire domain $M$ to functions with respect to individual points in the contour tree, and could be computed efficiently by sweeping through the tree, one regular node (mesh vertex) at a time.
Computation of the coefficients involved taking cells incident to each node, subtracting out their coefficients for the value range before the vertex was swept past and adding the coefficients for the value range after the vertex was swept past.

This work also showed that for a regular mesh, the regular node count on each superarc was a cheap and effective approximation of volume.
Pascucci~\cite{Pas01} showed how to compute the \emph{Betti numbers}~\cite{Pas01} for edges on the contour tree, which could potentially detect the characteristic toroidal configuration of vortex rings.

In recent years, efforts have focused on adding contour tree analysis to the principal topology toolkits: \code{TTK} (\textit{Topology Tool Kit})~\cite{TFL17} (now incorporated into \code{ParaView}), and \code{Viskores}~\cite{MSU16}. Either can be used for contour tree analysis, with \code{Viskores} performing better at scale, while \code{TTK} already supports irregular meshes.

Hristov~\etal~\cite{HWC20} demonstrated efficient parallel computation in \code{Viskores} for arbitrary contour tree properties using prefix-sums.
Their approach uses a variation on rake-and-contraction~\cite{MR85} called the \emph{hypersweep}, however, was applied only to compute volume by counting regular nodes.
In contrast, there is no explicit mechanism in \code{TTK} for computing properties through the contour tree sweep.

In summary, parallel contour tree algorithms exist, but computation of geometric properties on irregular meshes is incomplete.

\section{Problem Statement}
\label{sec:problemstatement}

In applying contour tree analysis to meteorological simulations of cloud formation, the obvious solution is to resample from the parcel-in-cell simulation to a rectilinear mesh, then run the existing tools, either in \code{TTK} or in \code{Viskores}. 
However, a problem arises - what is the appropriate sampling rate, and how much does it cost?

The parcel centroids are not uniformly distributed (\xref{sec:parcelincell}).
This means that sampling resolution on a regular grid must be driven by the closest pair of parcels.
Following the Nyquist sampling theorem~\cite{shannon1949communication}, we would need a resolution of no more than half of this distance.
For our PIC simulation, which has around $2,000,000$ parcels, this argues for a $4,096^3 = 68,719,476,736$ sampling grid resolution, and $>256GB$ of RAM just to store a single variable.

Previous experience has shown that the memory footprint of contour tree analysis is  $\approx200B$ or more per sample, implying the need for $> 10TB$ of RAM if the computation is performed on a single machine. At this point, we started asking if it was wiser to work directly with the PIC representation rather than re-sampling it.

Applying the knowledge derived from the previous work outlined above, we can see that the first step is to choose a suitable topology graph based on the parcel-in-cell simulations. Since these simulations already depend on tracking which parcels are near each other, the natural choice is to use the Delaunay triangulation, which we can then represent as an irregular mesh in \code{Viskores}.

Although all of the other steps are on principle solved, the remaining obstacle is the computation of volume, which is the property of interest to the meteorologists.  For an irregular mesh, such as a Delaunay triangulation, counting regular nodes is not an effective approximation, so \xref{sec:polynomialcoefficients} will develop the formulae necessary to compute volume correctly in an irregular mesh such as Delaunay.

Once we have done this, we will give the implementation details that were necessary to add this to the \code{Viskores} library in \xref{sec:implementation}, then compare the rectilinear and PIC-based results in \xref{sec:results}.

\section{Polynomial Coefficients}
\label{sec:polynomialcoefficients}

\newcommand{\Real}{\mathbb{R}}	
\newcommand{\Complex}{\mathbb{C}}	
\newcommand{\norm}[1]{\bigl\lVert#1\bigr\rVert}
\newcommand{\normvec}[1]{\bigl\lVert\vec{#1}\bigr\rVert}
\newcommand{\ktwo}{\frac{1}{h_{3}-h_{2}}}
\newcommand{\ttwo}{\frac{h-h_{2}}{h_{3}-h_{2}}}
\newcommand{\stwo}{\frac{h_3-h}{h_{3}-h_{2}}}

\newcommand{\Verts}[1][2ex]{\rule{1.7pt}{ #1 }}

We saw in \xref{sec:contourtreeanalysis} that area and volume are polynomial splines in the isovalue $h$~\cite{BPS97} and that the correct derivation needs Federer's co-area formula~\cite{federer_gmt_69} for area integration~\cite{zhou2018efficient}. 
\new{The polynomial must be expressed in standard form (such as $ah^2+bh+c$) for summing coefficients (e.g. $a, b, c$) with a hypersweep~\cite{HWC20} (\xref{sec:contourtreeanalysis}). 
Instead of expanding B-splines~\cite{zhou2018efficient} recursively into the standard form, we develop a direct geometric construction, making explicit how each coefficient arises from the tetrahedral mesh itself rather than treating them as abstract results of the spline formulation.}\\

\begin{figure}[H]
    \centering
    \includegraphics[width=1\linewidth]{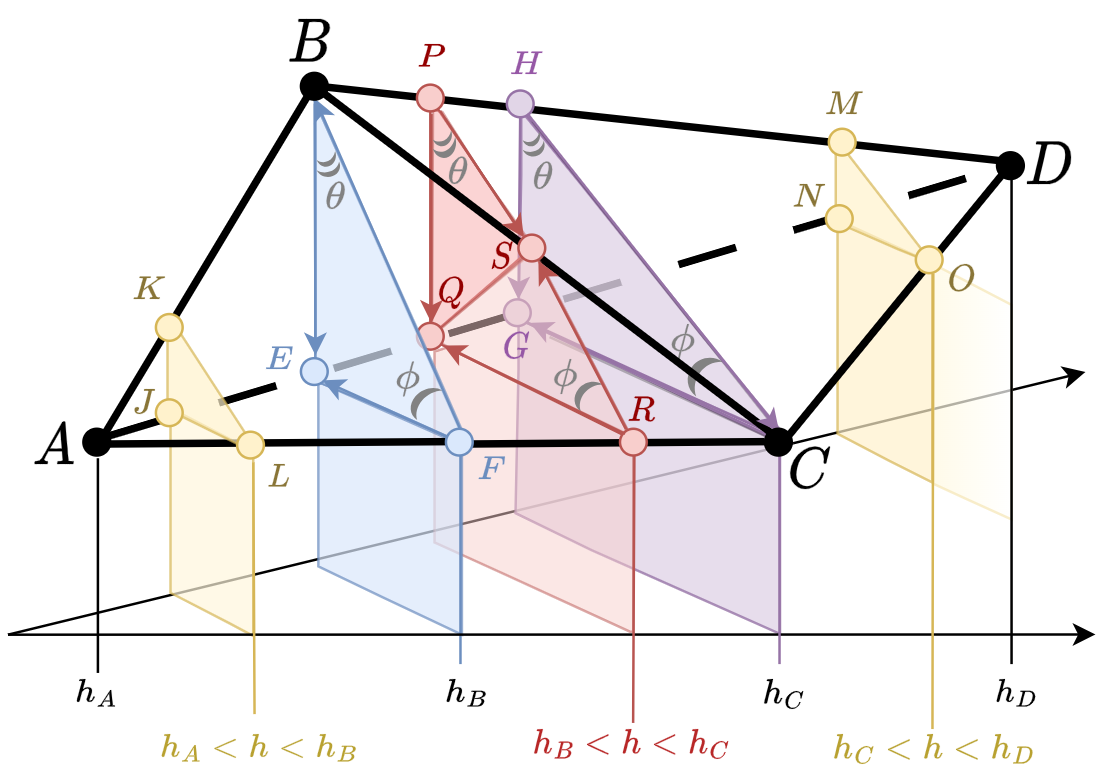}
    \caption{Tetrahedron $ABCD$ with values $h_A < h_B < h_C < h_D$.  Barycentric interpolation guarantees that all contours are planar and parallel, with constant angles $\theta,\phi$. Area at isovalue $h$ is computed with $\triangle JKL$, $\triangle PQS$ and $\triangle QRS$, or $\triangle MNO$, and volume is integrated from area as described in the text. }
    \label{fig:tetrahedron-labelled}
\end{figure}

Assume a tetrahedron has vertices $A,B,C,D$ with function values $h_A < h_B < h_C < h_D$ (\xref{fig:tetrahedron-labelled}). Barycentric contours in a tetrahedron are parallel planes.  Where contours meet tetrahedron edges, the angle is constant and is fixed by the contours intersecting with the edge dihedral. Vertices $E,F$ are defined by linear interpolation on edges $AD, AC$ by the triangular contour $BEF$ at $h = h_B = f(B)$. We get $Area(BEF)$ from half the cross product of $FB, FE$. We can then compute $\phi$, the angle at $F$ in $BEF$, or more importantly, $\sin (\phi)$.  Similarly, we compute vertices $G,H$ of contour $CGH$ at $h_C=f(C)$, $Area(CGH)$, and $\sin(\theta)$. The scalar triple product is used to compute $Volume(ABEF)$ and $Volume(DCGH)$.

We consider contours in the range $h_A < h < h_B$: in \xref{fig:tetrahedron-labelled}, the contour is a single triangle $JKL$, where $J, K, L$ are linearly interpolated along $AB, AE, AF$ by the parameter $r=\frac{h - h_A}{h_B - h_A}$. \\[0.05in]
It then follows  that: 
\begin{eqnarray*}
    Area(JKL)       &   =   &   r^2 Area(BEF)                                                             \nonumber   \\
                    &   =   &   \frac{Area(BEF)}{(h_B - h_A)^2} (h^2 - 2h_Ah +h_A^2)                      \nonumber   \\
    Volume(AJKL)    &   =   &   r^3 Volume(ABEF)                                                             \nonumber   \\
                    &   =   &   \frac{Volume(ABEF)}{(h_B - h_A)^3} (h^3 - 3h_Ah^2 + 3 h_A^2h - h_A^3)        \nonumber   \\
\end{eqnarray*}

For isovalues in the range $h_C < h < h_D$, the position is symmetric, and we parameterize on $t=\frac{h_D - h}{h_D - h_C}$ to get:
\begin{eqnarray*}
    Area(MNO)       &   =   &   \frac{Area(CGH)}{(h_D - h_C)^2} (h^2 - 2h_Dh +h_D^2)                      \nonumber   \\
    Volume(DMNO)    &   =   &   \frac{Volume(DCGH)}{(h_D - h_C)^3} (h_4^3 - 3h\cdot h_D^2 + 3 h^2h_4 - h^3)      \nonumber   \\
\end{eqnarray*}

However, since we are actually interested in the volume to the left of $\triangle MNO$, we convert the volume by subtraction:
\begin{equation*}
    Volume(ABCMNO)  =   Volume(ABCD) - Volume(DMNO)                                                 \nonumber   \\
\end{equation*}

This leaves the middle section, with isovalues $h_B < h < h_C$.  We know $Volume(ABEF)$ and want the $Volume(BEFPQRS)$. We consider the area of quad $PQRS$, noting that angles $\theta, \phi$ are fixed by the edge dihedral, and using $PQ = \left|PQ\right|,PS=\left|PS\right|$, and so forth. Now, we know that $P$ and $Q$ are linearly interpolated with parameter $s=\frac{h - h_B}{h_C - h_B}$ and $1-s = \frac{h_C - h}{h_C - h_B} $ between $B$ and $H$ respectively, so:
\begin{eqnarray*}
 	PQ    	&   =   &   s \cdot HG + (1-s) BE         								\nonumber   \\
	PS		&	=	&	s \cdot HC										\nonumber	
\end{eqnarray*}
\begin{eqnarray*}
	PQ \cdot PS	&	=	&	 \left(\frac{HG - BE}{h_C - h_B} h + \frac{h_C BE - h_B  HG}{h_C - h_B}\right) 								\nonumber \\
				&		&	\qquad\qquad\cdot \left(\frac{HC}{h_C - h_B} h + \frac{- h_B  HC }{h_C - h_B}\right)						\nonumber	\\
				&	=	&	\frac{ \left(HG - BE\right)  HC}{(h_C - h_B)^2} h^2 														\nonumber	\\
				&		&	+ \frac{ 2 h_B HG\cdot HC + (h_C - h_B) BE \cdot HC}{(h_C - h_B)^2} h 								\nonumber	\\
				&		&	+ \frac{ \left(h_B^2 HG - h_B h_C BE \right) HC }{(h_C - h_B)^2}  											\nonumber	\\
\end{eqnarray*}

Repeating the process for $\triangle QRS$ gives us a similar set of terms:
\begin{eqnarray*}
 	RQ    	&   =   	&   s \cdot EF + (1-s) GC         							\nonumber   \\
	RS		&	=	&	(1-s) \cdot BF									\nonumber	\\
	RQ \cdot RS	&	=	&	 \left(\frac{EF - GC}{h_C - h_B} h + \frac{h_C GC - h_B  EF}{h_C - h_B}\right) 							\nonumber \\
				&		&	\qquad\qquad\cdot \left(\frac{BF}{h_C - h_B} h_C - \frac{BF}{h_C - h_B} h\right)						\nonumber	\\
				&	=	&	\frac{ \left(GC - EF\right) \cdot(-BF)}{(h_C - h_B)^2} h^2 								\nonumber	\\
				&		&	+ \frac{-2 h_C BF\cdot EF + (h_C + h_B) GC \cdot BF}{(h_C - h_B)^2} h 								\nonumber	\\
				&		&	+ \frac{ \left(h_C^2 EF - h_B h_C GC \right) BF }{(h_C - h_B)^2}  											\nonumber	\\
\end{eqnarray*}

Finally, we use these to compute:
\begin{eqnarray*}
	Area(PQRS)(h)) 	&	=	&	Area(\triangle PQS) + Area(\triangle QRS)		\nonumber	\\
				&	=	&	\frac{1}{2}\sin(\theta) PQ \cdot PS + \frac{1}{2}\sin(\phi) RQ \cdot RS		\nonumber	\\
				&	=	&	\alpha h^2 + \beta h + \gamma		\nonumber	\\
\end{eqnarray*}
\begin{eqnarray*}
	\alpha		&	=	&	\frac{\sin(\theta) \left(HG - BE\right)  HC - \sin(\phi) \left(GC - EF\right)  BF}{2 (h_C - h_B)^2}		\nonumber	\\
	\beta		&	=	&	\frac{\sin(\theta)\left(2 h_B HG\cdot HC + (h_C - h_B) BE \cdot HC\right) }{2 (h_C - h_B)^2} 		\nonumber	\\
				&		&	+ \frac{\sin(\phi) \left(-2 h_C BF\cdot EF + (h_C + h_B) GC \cdot BF\right)}{2 (h_C - h_B)^2}		\nonumber	\\
	\gamma		&	=	&	\frac{\sin(\theta) \left(h_B^2 HG - h_B h_C BE \right) HC }{2 (h_C - h_B)^2} 		\nonumber	\\
				&		&	+ \frac{ \sin(\phi) \left(h_C^2 EF - h_B h_C GC \right) BF}{2 (h_C - h_B)^2}		\nonumber	\\	
\end{eqnarray*}

For volume, we apply Federer's co-area formula~\cite{federer_gmt_69} and integrate with respect to $h$.  This computes the interval volume between two isosurfaces by applying a correction factor of the inverse of the gradient magnitude~\cite{DCM12}.  In a barycentric interpolant, gradient magnitude is constant, and is the difference between isovalues $h_B, h_C$  divided by the perpendicular distance between $BEF$ and $CGH$, using the normal form of plane $BEF$, $\vec{n} = \frac{EF \times BF}{|EF \times BF|} $ and points $B,H$, i.e. $\delta = \vec{n}\cdot B - \vec{n} \cdot H $. Since $\delta$ is independent of $h$, it can be moved outside the integral. The volume polynomial for $BEFPQRS$ is then:
\begin{equation*}
    Volume(BEFPQRS)(h) = \frac{\alpha \delta}{3} h^3 + \frac{\beta \delta}{2} h^2 + \gamma \delta h + D \nonumber  
\end{equation*}
where $D$, the constant of integration can be determined by observing that for $h = h_B$, the volume must be $0$, since we are interested in the $\Delta$volume (change since the last vertex), and back-substituting.

Once we have the correct coefficients, we can compute local (per-tetrahedron) coefficient changes (deltas) in a parallel pre-processing step. Then, these are summed efficiently in the polynomial hypersweep (\xref{sec:polynomialhypersweep}) to obtain the volume for each superarc.

\begin{figure*}
	\centering
        \includegraphics[height=75mm, alt={10kprunedcontourtree}]{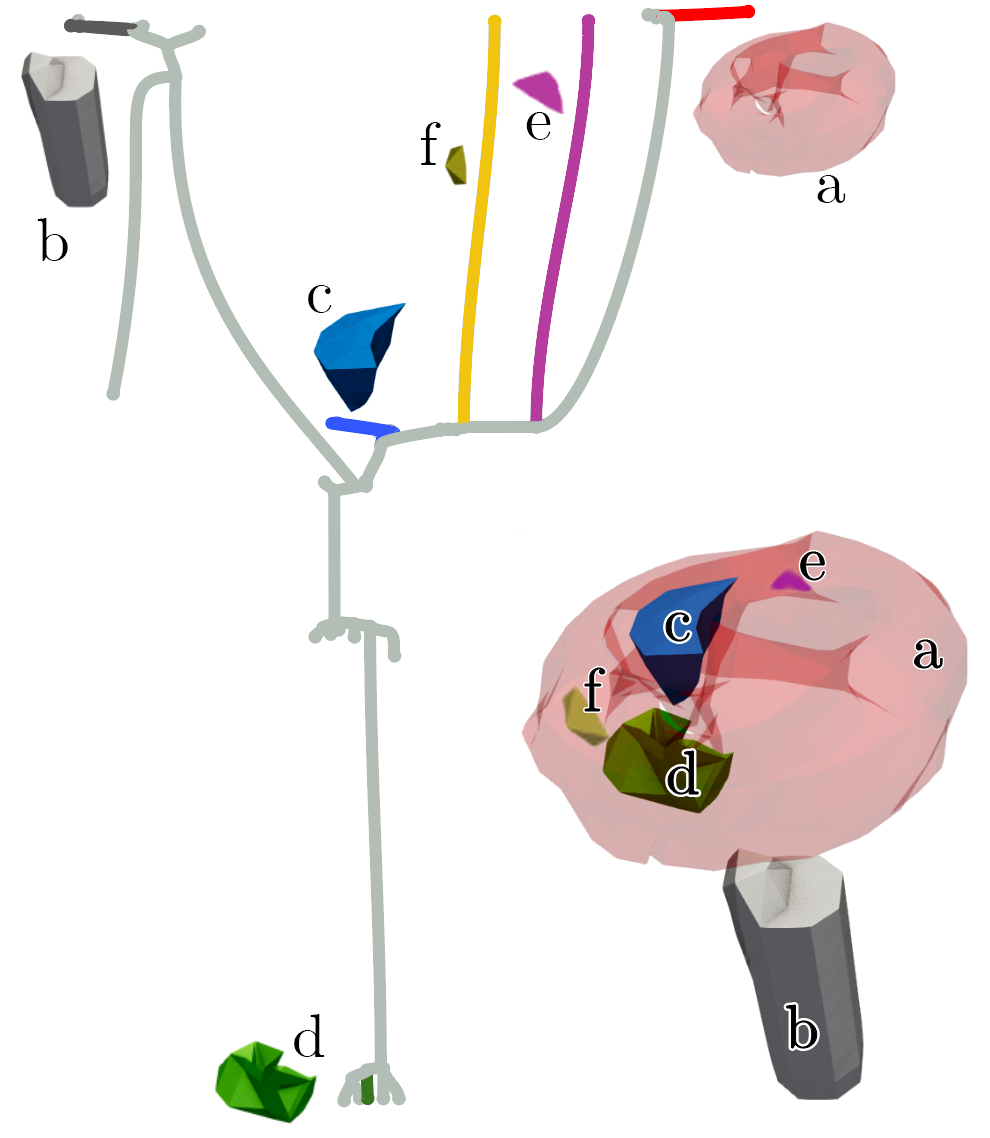}
        \includegraphics[height=75mm, alt={Resampled 64cubed}]{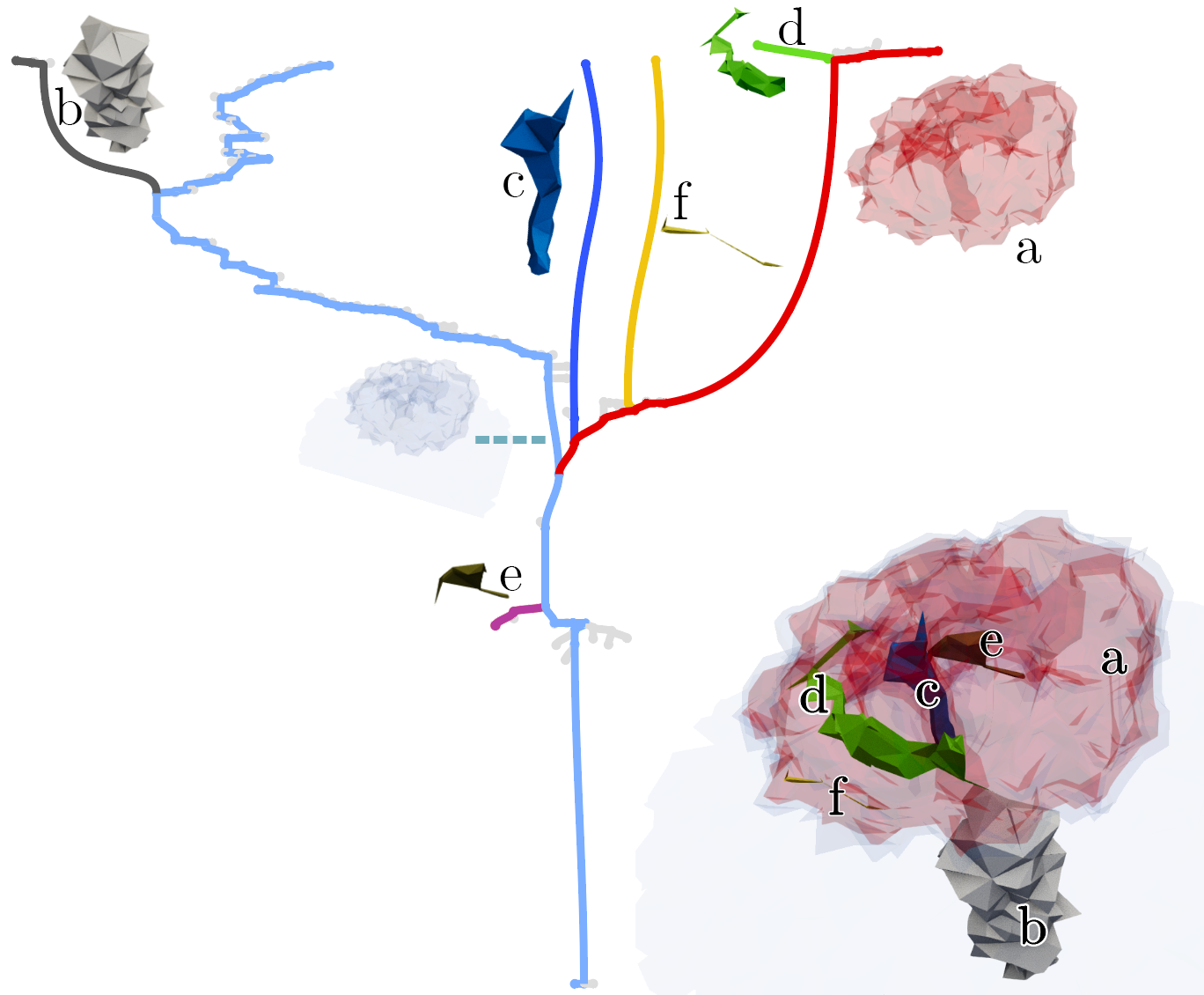}
    \caption{Contour trees computed from the $24^3$ grid (left) and $10K$ subsampled particles (right).
For each major branch (ranked by volume), flexible isosurface visualizations are shown and color-coded, alongside the full datasets.
Light-gray branches denote low-volume features.}
    \label{fig:ContoursVsTree}
\end{figure*}

\section{Polynomial Hypersweep}
\label{sec:polynomialhypersweep}

We saw in \xref{sec:contourtreeanalysis} that the existing pipeline was implemented for regular grids, and that volume was approximated by the regular node count.  In \xref{sec:polynomialcoefficients}, we developed the correct polynomial coefficients. 

Since \code{Viskores} is based on the PRAM model, we consider how to extend  hypersweeps~\cite{HWC20} to update the polynomial coefficients. For each tetrahedral cell $K$ incident to vertex $v$, the sweep subtracts coefficients $k_1,k_2, \ldots, k_n$ before the vertex is swept, then adds coefficients $K_1,K_2, \ldots, K_n$ afterwards.  This is equivalent to adding $\delta_1 = K_1 - k_1, \ldots, \delta_n = K_n - k_n$, which can be computed independently for each vertex in each cell.

The solution is then to compute the deltas for each cell in the mesh at each vertex (wlog, in the downwards direction).  Assign the deltas to the vertex, and sum all deltas for each vertex to produce a single set of deltas per vertex.  The correct coefficients in each regular arc can then be computed as a prefix sum using a hypersweep.

This is not yet complete, as hypersweeps may reverse the sweep direction.  The existing code subtracts node count from total count for the entire data set, then adds one because the origin supernode of each superarc is counted as part of the superarc.  For polynomial coefficients, it is necessary to modify the approach so that deltas are stored at each supernode as well as at each superarc.

\section{Implementation}
\label{sec:implementation}

Given the polynomial coefficients in \xref{sec:polynomialcoefficients}, we can now turn to the implementation details for computing PIC-based contour trees.

The key decision was whether to implement the new exact volume simplification pipeline in \code{VTK-m} (now called \code{Viskores}) or \code{TTK}. 
As in \xref{sec:contourtreeanalysis}, \code{TTK} already has irregular meshes implemented, but does not have the hypersweep framework needed for summing the coefficients described in \xref{sec:polynomialcoefficients}.
\code{TTK} is integrated to \code{ParaView}, while \code{Viskores} form the basis of distributed contour tree computation methods and is faster at large scales \cite{CWR22}. 
We therefore decided to implement the pipeline in \code{Viskores}, for which there are five principal changes required:
\begin{itemize}
	\item	Delaunay tetrahedralization 
	\item	Replacing grid-based code with mesh-based code	
	\item	Implementing polynomial coefficient computation
	\item	Modifications to hypersweep
	\item	Visualization pipeline changes: flexible isosurface adaptation
\end{itemize}


\paragraph{Delaunay Tetrahedralization:} Since Delaunay meshing is not yet in the \code{Viskores} codebase, we computed a Delaunay  tetrahedralization from the parcel centroids externally with \code{TetGen}~\cite{hang2015tetgen}, as shown in \xref{fig:teaser}(c), then read the resulting mesh into \code{Viskores}.

\paragraph{Grid To Mesh:} Given prior experience with the contour tree, considerable effort was expended in the original implementation of \code{Viskores} to ensure that the contour tree algorithm was abstracted rather than built directly on the grid. 
It was previously implemented by building a \code{Mesh} class on top of the input data, for 2D and 3D \emph{Freudenthal} simplicial meshes, plus the ability to compute contour trees based on \emph{Marching Cubes} connectivity.  
Since the contour tree is a valid \emph{topology graph} for computing itself, a \code{Mesh} class could effectively encapsulate a contour tree as input to the computation, permitting an easy early way to distributed computation.

In all cases, access to the \code{Mesh} was abstracted, but in two places, there was an assumption of vertex degree less than $32$ for all vertices, allowing the use of a bitmask for efficient representation of vertex neighborhoods. 
This same vertex bound allowed us to run worklets that iterated around each vertex in early stages: for arbitrary-degree vertices, this can degenerate to serial performance.

We therefore implemented a new \code{Mesh} class to encapsulate an arbitrary \code{TopologyGraph}, and rewrote the section of code that iterated around a vertex to substitute a PRAM-efficient segmented sort to select the best ascending edge for each vertex. 
Since the original code was templated on the \code{Mesh}, it was not difficult to ensure that the existing code path was preserved for grid-based data.

\paragraph{Polynomial Coefficients:} As seen in \xref{sec:polynomialcoefficients}, further code was also required to compute the correct polynomial coefficients and deltas as pre-processing. 
This was done with a single parallel loop to compute all of the deltas for each cell, then a segmented sort to collect them in segments corresponding to each vertex in the mesh. 
Finally, a segmented prefix sum computed the total delta at each vertex, which was used as input to the following hypersweep. 

\paragraph{Modifications to Hypersweep:} The existing hypersweep code assumed a single (integer) value per vertex, and inversion could be easily performed by subtracting from the total volume and adding $1$. 
For full polynomial coefficients, we had to upgrade the Hypersweep to a version templated on a data type (in our case, vectors of four doubles to represent the coefficients) and which correctly handled arbitrary deltas at each supernode.

\paragraph{Visualization Pipeline:}	Previous work~\cite{HWC20} computed the branch decomposition based on node counts, and had to be templated to allow arbitrary data types.  We also carried over the flexible isosurface~\cite{CSV10} extraction of single contours by first reconstructing entire surfaces at critical points with marching tetrahedra, then choosing only those triangles that mapped to the correct superarc, as before. This is different for the main branch (the one with the largest volume). As it connects to many other branches, any extracted isosurface would either occlude inner structures or be occluded by them. Therefore, we select the middle iso-value, and later use transparency to show it.\\

As we will see in \xref{sec:results}, these modifications were successful, and we expect to contribute them to \code{Viskores} in due course.

\section{Results}
\label{sec:results}
We used an EPIC~\cite{frey20233d} simulation of a developing cloud at a mature stage, with $\sim$2 million parcels, and its resampled version onto a $1024^3$ grid. We evaluated performance, accuracy, and quality, comparing the time cost and memory use of our code to contour tree implementations in \code{Viskores} and \code{TTK}. We then examined the contour tree structure and displayed flexible isosurfaces side-by-side.

\paragraph{Hardware:} Because contour tree computation has a significant memory footprint (\xref{sec:problemstatement}), we chose to run our benchmarks on the Aire HPC Facility. We use the standard compute partition that offers CPU nodes with AMD Dual 84 core 2.2GHz (9634 Genoa-X) processors and 768 GB DDR5-4800 memory. 

\new{
\paragraph{Software:} Our method builds on \code{VTK-m} $2.3.0$ (at the time of its rebranding as \code{Viskores} $1.0.0$).
We compare our \code{Viskores} irregular mesh implementation against \code{TTK} $1.3.0$ (the most recent stable release at the time of our evaluation), and include three contour tree construction baselines: on regular grids (timing, memory \xref{tab:grids-vtkm-ttk}), and on irregular meshes (timing \xref{tab:irregular-vtkm-ttk-timing}, memory \xref{tab:irregular-vtkm-ttk-memory}).
All benchmarks exclude I/O and were run in parallel, on 128 threads.

}

\paragraph{Data Sets:} We use a parcel-based simulation generated by atmospheric scientists for studying cloud formation \cite{frey20233d}. The simulation tracks two million parcels and we use the parcel centroids from one of the time steps when the cloud has already matured. We use the \textit{total water specific humidity} scalar field (i.e. \textit{kg} of water vapour plus liquid water per \textit{kg} of air), as the scientists are interested in studying the cloud droplet distribution (\xref{sec:meteorology}).

\begin{table*}[]
	\centering
	\begin{tabular}{|cc||c|ccc|c|c||c|c|c|c|c|}
        \hline
	
	\multicolumn{2}{|c||}{} & 
	\multicolumn{6}{c||}{\textbf{Timing}} & 
	\multicolumn{5}{c|}{\textbf{Memory Usage}} \\ 
    
	\multicolumn{2}{|c||}{\textbf{Input}} & 
	\multicolumn{4}{c|}{\textbf{\cellcolor[HTML]{F2F2F2}Viskores (VTK-m)}} & 
	\multicolumn{1}{c|}{\textbf{\cellcolor[HTML]{F2F2F2}TTK}} & 
	\multicolumn{1}{c||}{\cellcolor[HTML]{F2F2F2}} & 
	\multicolumn{3}{c|}{\textbf{\cellcolor[HTML]{F2F2F2}Viskores (VTK-m)}} & 
	\multicolumn{2}{c|}{\textbf{\cellcolor[HTML]{F2F2F2}TTK}} 
        \\ \cline{4-7}\cline{9-13} 
	
	\multicolumn{2}{|c||}{} & 
	\multicolumn{4}{c|}{\cellcolor[HTML]{D9D9D9}... \textit{of which}} &  
	\multicolumn{1}{c|}{\cellcolor[HTML]{D9D9D9}} & 
	\multicolumn{1}{c||}{\cellcolor[HTML]{F2F2F2}} & 
	\multicolumn{2}{c|}{\cellcolor[HTML]{D9D9D9}... \textit{of which}} & 
	\multicolumn{1}{c|}{\cellcolor[HTML]{F2F2F2}} & 
	\multicolumn{1}{c|}{\cellcolor[HTML]{D9D9D9}} & 
	\multicolumn{1}{c|}{\cellcolor[HTML]{F2F2F2}} \\ \cline{1-2}\cline{4-6}\cline{10-10} 
	
	\rotatebox{90}{Resolution\textcolor{white}{,}}\rotatebox{90}{\textcolor{gray}{{(cubed)}}}			&
	\rotatebox{90}{Input Size}							 		&
	\rotatebox{90}{\cellcolor[HTML]{D9D9D9}Total Time\textcolor[HTML]{D9D9D9}{,}}\rotatebox{90}{\textcolor{gray}{ (seconds)}}			&
   	\rotatebox{90}{Construction}\rotatebox{90}				&
   	\rotatebox{90}{Weights}\rotatebox{90}							 		&
	\rotatebox{90}{Branch}\rotatebox{90}{Decomposition}\rotatebox{90}							 		&
	\rotatebox{90}{\cellcolor[HTML]{D9D9D9}Total Time\textcolor[HTML]{D9D9D9}{,}}\rotatebox{90}{\textcolor{gray}{ (seconds)}}			&
   	\rotatebox{90}{\cellcolor[HTML]{F2F2F2}Time Ratio\textcolor[HTML]{D9D9D9}{,}}\rotatebox{90}{\textcolor{gray}{({\code{Viskores} / \code{TTK}})}}			&
   	\rotatebox{90}{\cellcolor[HTML]{D9D9D9}Peak Memory}\rotatebox{90}{\textcolor{gray}{{(MiB)}}}			&
   	\rotatebox{90}{Raw Data\textcolor{white}{,}}\rotatebox{90}{\textcolor{gray}{{(MiB)}}}				&
   	{\centering \raisebox{0.85\height}{\parbox{0.79cm}{\cellcolor[HTML]{F2F2F2}Bytes\\ per\\ Data \\Point}}}										&
   	\rotatebox{90}{\cellcolor[HTML]{D9D9D9}Peak Memory}\rotatebox{90}{\textcolor{gray}{{(MiB)}}}			&
   	{\centering \raisebox{0.85\height}{\parbox{0.79cm}{\cellcolor[HTML]{F2F2F2}Bytes\\ per\\ Data \\Point}}}										\\
   	\cline{1-2}\cline{4-6}\cline{10-10} 
24   & 14K  & \cellcolor[HTML]{D9D9D9}0.068  & 0.0662               & 0.0003 & 0.0019 & \cellcolor[HTML]{D9D9D9}0.003   & \cellcolor[HTML]{F2F2F2}22.08 & \cellcolor[HTML]{D9D9D9}1.6    & 0.053    & \cellcolor[HTML]{F2F2F2}111       & \cellcolor[HTML]{D9D9D9}3.232   & \cellcolor[HTML]{F2F2F2}234                  \\
32   & 32K  & \cellcolor[HTML]{D9D9D9}0.073  & 0.0708               & 0.0005 & 0.0021 & \cellcolor[HTML]{D9D9D9}0.006   & \cellcolor[HTML]{F2F2F2}11.80 & \cellcolor[HTML]{D9D9D9}3.48   & 0.125    & \cellcolor[HTML]{F2F2F2}102       & \cellcolor[HTML]{D9D9D9}6.713   & \cellcolor[HTML]{F2F2F2}205                  \\
48   & 110K & \cellcolor[HTML]{D9D9D9}0.120  & 0.1162               & 0.0011 & 0.0024 & \cellcolor[HTML]{D9D9D9}0.021   & \cellcolor[HTML]{F2F2F2}5.536 & \cellcolor[HTML]{D9D9D9}11.19  & 0.422    & \cellcolor[HTML]{F2F2F2}97        & \cellcolor[HTML]{D9D9D9}21.07   & \cellcolor[HTML]{F2F2F2}191                  \\
64   & 262K & \cellcolor[HTML]{D9D9D9}0.137  & 0.1319               & 0.0022 & 0.0028 & \cellcolor[HTML]{D9D9D9}0.040   & \cellcolor[HTML]{F2F2F2}3.298 & \cellcolor[HTML]{D9D9D9}26.29  & 1        & \cellcolor[HTML]{F2F2F2}97        & \cellcolor[HTML]{D9D9D9}48.93   & \cellcolor[HTML]{F2F2F2}187                  \\
96   & 880K & \cellcolor[HTML]{D9D9D9}0.167  & 0.1571               & 0.0066 & 0.0031 & \cellcolor[HTML]{D9D9D9}0.120   & \cellcolor[HTML]{F2F2F2}1.310 & \cellcolor[HTML]{D9D9D9}88.09  & 3.375    & \cellcolor[HTML]{F2F2F2}96        & \cellcolor[HTML]{D9D9D9}163.8   & \cellcolor[HTML]{F2F2F2}185                  \\
128  & 2.1M & \cellcolor[HTML]{D9D9D9}0.241  & 0.2226               & 0.0151 & 0.0034 & \cellcolor[HTML]{D9D9D9}0.254   & \cellcolor[HTML]{F2F2F2}0.877 & \cellcolor[HTML]{D9D9D9}208.5  & 8        & \cellcolor[HTML]{F2F2F2}95        & \cellcolor[HTML]{D9D9D9}395.3   & \cellcolor[HTML]{F2F2F2}188                  \\
192  & 7M   & \cellcolor[HTML]{D9D9D9}0.482  & 0.4244               & 0.0510 & 0.0071 & \cellcolor[HTML]{D9D9D9}0.903   & \cellcolor[HTML]{F2F2F2}0.470 & \cellcolor[HTML]{D9D9D9}703.3  & 27       & \cellcolor[HTML]{F2F2F2}96        & \cellcolor[HTML]{D9D9D9}1,296   & \cellcolor[HTML]{F2F2F2}183                  \\
256  & 16M  & \cellcolor[HTML]{D9D9D9}0.895  & 0.7636               & 0.1204 & 0.0106 & \cellcolor[HTML]{D9D9D9}2.623   & \cellcolor[HTML]{F2F2F2}0.291 & \cellcolor[HTML]{D9D9D9}1,628  & 64       & \cellcolor[HTML]{F2F2F2}93        & \cellcolor[HTML]{D9D9D9}3,103   & \cellcolor[HTML]{F2F2F2}185                  \\
384  & 57M  & \cellcolor[HTML]{D9D9D9}2.285  & 1.8687               & 0.4032 & 0.0132 & \cellcolor[HTML]{D9D9D9}11.730  & \cellcolor[HTML]{F2F2F2}0.159 & \cellcolor[HTML]{D9D9D9}5,489  & 216      & \cellcolor[HTML]{F2F2F2}93        & \cellcolor[HTML]{D9D9D9}5,953   & \cellcolor[HTML]{F2F2F2}105                  \\
512  & 134M & \cellcolor[HTML]{D9D9D9}4.674  & 3.5602               & 0.9544 & 0.1590 & \cellcolor[HTML]{D9D9D9}34.874  & \cellcolor[HTML]{F2F2F2}0.102 & \cellcolor[HTML]{D9D9D9}13,010 & 512      & \cellcolor[HTML]{F2F2F2}93        & \cellcolor[HTML]{D9D9D9}14,030  & \cellcolor[HTML]{F2F2F2}105                  \\
1024 & 1B   & \cellcolor[HTML]{D9D9D9}37.574 & 29.5676              & 7.7865 & 0.2195 & \cellcolor[HTML]{D9D9D9}504.800 & \cellcolor[HTML]{F2F2F2}0.059 & \cellcolor[HTML]{D9D9D9}96,700 & 4096     & \cellcolor[HTML]{F2F2F2}86        & \cellcolor[HTML]{D9D9D9}108,594 & \cellcolor[HTML]{F2F2F2}101                  \\
\hline
\end{tabular}
    \caption{Comparison of \textit{timings and memory usage} for \code{Viskores} (\code{VTK-m}) and \code{TTK} for grid-based data. We also note that it takes $\sim$23 minutes just to grid the original simulation onto the $1024^3$ mesh. The resolutions $24^3$ to $512^3$ are down-sampled versions of the $1024^3$ grid.}
    \label{tab:grids-vtkm-ttk}
\end{table*}


\begin{table*}[]
	\centering
	\begin{tabular}{|c|c|c|cc|c|cc||c|cc|c|}
	\cline{1-12}

        \multicolumn{1}{|c|}{} & 
	\multicolumn{1}{c|}{Pre-} & 
	\multicolumn{6}{c||}{\textbf{Viskores (VTK-m)}} & 
	\multicolumn{3}{c|}{\textbf{TTK}} &
	\multicolumn{1}{c|}{Ratio}\\ 
	

        \multicolumn{1}{|c|}{} & 
	\multicolumn{1}{c|}{Processing} & 
	\multicolumn{3}{c|}{\cellcolor[HTML]{D9D9D9}... \textit{of which}} &  
	\multicolumn{3}{c||}{\cellcolor[HTML]{F2F2F2} ... \textit{of which}}& 
	\multicolumn{3}{c|}{\cellcolor[HTML]{D9D9D9}... \textit{of which}} & 
	\multicolumn{1}{c|}{\cellcolor[HTML]{F2F2F2}} 
    
    \\ \cline{2-2}\cline{4-5}\cline{7-8}\cline{10-11}
	
	\rotatebox{90}{Num. of Points}			&
	\rotatebox{90}{Delaunay}\rotatebox{90}{\textcolor{gray}{(seconds)}}							 		&
	   	\rotatebox{90}{\cellcolor[HTML]{D9D9D9}Construction\textcolor[HTML]{D9D9D9}{,}\textcolor[HTML]{D9D9D9}{,}}\rotatebox{90}{Total Time\textcolor[HTML]{D9D9D9}{,} }\rotatebox{90}{{\textcolor{gray}{(seconds)}}}			&
   	\rotatebox{90}{Traversal}\rotatebox{90}				&
   	\rotatebox{90}{\texttt{CTA}\textcolor{white}{,}}\rotatebox{90}	{Worklet}\rotatebox{90}							 		&
       	\rotatebox{90}{\cellcolor[HTML]{F2F2F2}Simplfication\textcolor[HTML]{F2F2F2}{,}\textcolor[HTML]{F2F2F2}{,}}\rotatebox{90}{Total Time\textcolor[HTML]{F2F2F2}{,} }\rotatebox{90}{{\textcolor{gray}{(seconds)}}}			&

	\rotatebox{90}{Volume\textcolor{white}{,}}\rotatebox{90}{Weights}\rotatebox{90}									&
   	\rotatebox{90}{Branch\textcolor{white}{,}}\rotatebox{90}{Decomposition}\rotatebox{90}			&
   	\rotatebox{90}{\cellcolor[HTML]{D9D9D9}Construction\textcolor[HTML]{D9D9D9}{,}\textcolor[HTML]{D9D9D9}{,}}\rotatebox{90}{Total Time\textcolor[HTML]{D9D9D9}{,} }\rotatebox{90}{{\textcolor{gray}{(seconds)}}}			&
   	\rotatebox{90}{Traversal\textcolor{white}{,}}\rotatebox{90}{\textcolor{gray}{{(\texttt{OneSkeleton})}}}				&
   	\rotatebox{90}{\texttt{FTMTree}\textcolor{white}{,}}\rotatebox{90}{Filter}				&
   	\rotatebox{90}{\cellcolor[HTML]{F2F2F2}Time Ratio\textcolor{white}{,}}\rotatebox{90}{\textcolor{gray}{(\texttt{CTA /FTMTree})}}			
    
    \\ \cline{1-2}\cline{4-5} \cline{7-8} \cline{10-11} 
    
10K  & 0.040   & \cellcolor[HTML]{D9D9D9}0.142  & 0.022 & 0.120   & \cellcolor[HTML]{F2F2F2}0.020   & 0.014   & 0.006  & \cellcolor[HTML]{D9D9D9}0.019    & 0.011  & 0.008   & \cellcolor[HTML]{F2F2F2}15.01 \\
30K  & 0.168   & \cellcolor[HTML]{D9D9D9}0.210  & 0.055 & 0.155   & \cellcolor[HTML]{F2F2F2}0.044   & 0.037   & 0.007  & \cellcolor[HTML]{D9D9D9}0.045    & 0.033  & 0.012   & \cellcolor[HTML]{F2F2F2}12.93 \\
100K & 0.425   & \cellcolor[HTML]{D9D9D9}0.329  & 0.079 & 0.250   & \cellcolor[HTML]{F2F2F2}0.113   & 0.105   & 0.007  & \cellcolor[HTML]{D9D9D9}0.125    & 0.098  & 0.027   & \cellcolor[HTML]{F2F2F2}9.24  \\
250K & 1.455   & \cellcolor[HTML]{D9D9D9}0.499  & 0.156 & 0.343   & \cellcolor[HTML]{F2F2F2}0.238   & 0.230   & 0.008  & \cellcolor[HTML]{D9D9D9}0.312    & 0.256  & 0.056   & \cellcolor[HTML]{F2F2F2}6.13  \\
1M   & 5.470   & \cellcolor[HTML]{D9D9D9}1.214  & 0.331 & 0.883   & \cellcolor[HTML]{F2F2F2}1.049   & 1.038   & 0.011  & \cellcolor[HTML]{D9D9D9}1.245    & 1.046  & 0.199   & \cellcolor[HTML]{F2F2F2}4.44  \\
2.1M & 13.889  & \cellcolor[HTML]{D9D9D9}2.075  & 0.646 & 1.429   & \cellcolor[HTML]{F2F2F2}2.155   & 2.140   & 0.016  & \cellcolor[HTML]{D9D9D9}2.603    & 2.162  & 0.441   & \cellcolor[HTML]{F2F2F2}3.24  \\
30M* & 225.84* & \cellcolor[HTML]{D9D9D9}42.39* & 7.72* & 34.671* & \cellcolor[HTML]{F2F2F2}71.758* & 70.998* & 0.759* & \cellcolor[HTML]{D9D9D9}102.60*  & 36.86* & 65.743* & \cellcolor[HTML]{F2F2F2}0.53* \\

\cline{1-12}
\end{tabular}
    \caption{Comparison of contour tree construction \textit{timings} on an irregular mesh in \code{Viskores} (\code{VTK-m}) and \code{TTK}.\\  Corresponding filters are: \texttt{CTA} (\texttt{ContourTreeAugmented}) in \texttt{Viskores}, and \texttt{FTMTree} (\textit{Fibonacci Task-based Merge tree)} in \texttt{TTK}.\\ Row marked with '*' comes from a different, 30M parcel simulation for time estimation on larger data.}
    \label{tab:irregular-vtkm-ttk-timing}
\end{table*}

\begin{figure*}[t]
    \centering
    \subfigure[Time usage trend comparison. Data from \xref{tab:grids-vtkm-ttk}, \xref{tab:irregular-vtkm-ttk-timing}]{
        \includegraphics[height=65mm]{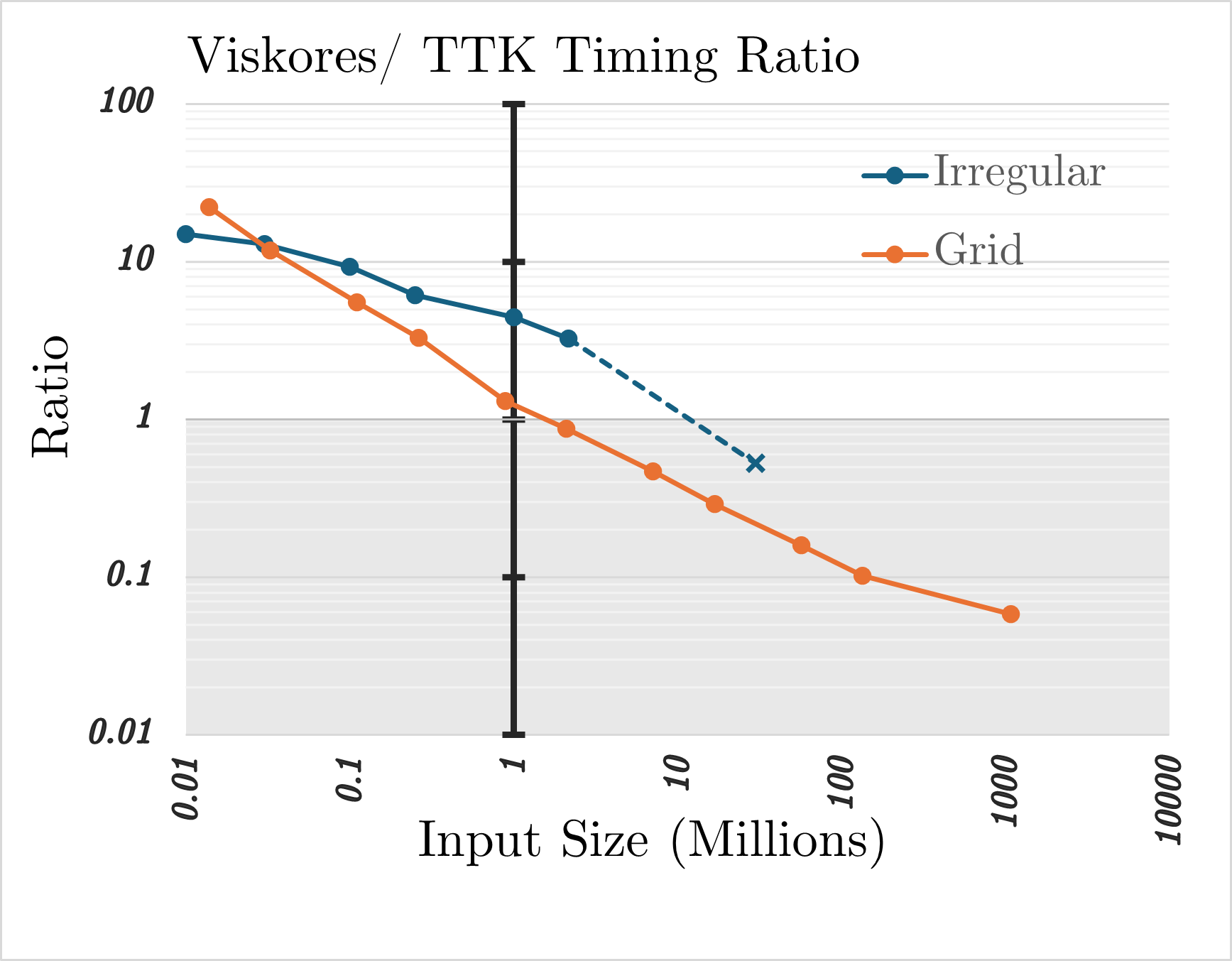}
        \label{fig:time-log-plot}
    }
    \subfigure[Memory usage trend comparison. Data from \xref{tab:grids-vtkm-ttk}, \xref{tab:irregular-vtkm-ttk-memory}]{
        \includegraphics[height=65mm]{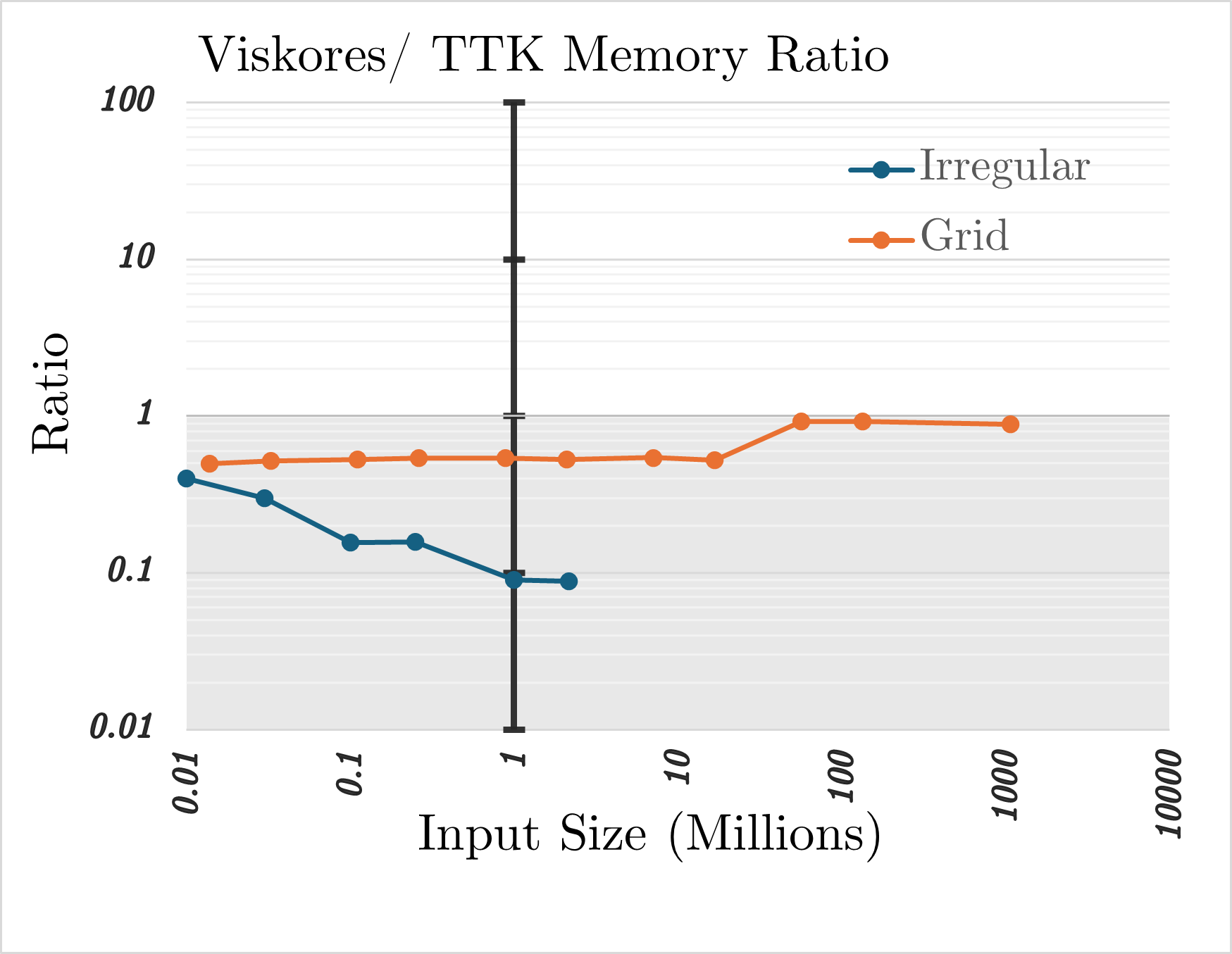}
        \label{fig:memory-log-plot}
    }

    \caption{Logarithmic plots showing the $\frac{\code{Viskores}}{\code{TTK}}$ ratios in contour tree construction \textit{timings} and \textit{memory use}. \\The x-axis shows the input sizes (millions of vertices); y-axis shows the ratios, with the shaded region below $y=1$ indicating better \code{Viskores} performance. Blue lines are measures on irregular data: (a) \xref{tab:irregular-vtkm-ttk-timing} (Time Ratio \texttt{CTA/FTMTree}), (b) \xref{tab:irregular-vtkm-ttk-memory} (Bytes per Data Point: \code{Viskores} (Construction) / \code{TTK}). Dashed blue line with the cross marker in (a) is from a larger, different (30M) simulation. Orange lines indicate grid data measures: (a) \xref{tab:grids-vtkm-ttk} (Time Ratio \texttt{CTA/FTMTree}), (b) \xref{tab:grids-vtkm-ttk} (Bytes per Data Point: \code{Viskores} / \code{TTK}).}
    \label{fig:logplots}
\end{figure*}

\begin{table*}[]
	\centering
	\begin{tabular}{|cc|c|ccc|cc|c||c|c|c|}
	\cline{1-12}
	
	\multicolumn{2}{|c|}{\textbf{Input}} & 
	\multicolumn{7}{c||}{\textbf{Viskores (VTK-m)}} & 
	\multicolumn{2}{c|}{\textbf{TTK}} &
        \multicolumn{1}{c|}{\textbf{Ratio}}
        \\ \cline{3-12}
	
	
	\multicolumn{2}{|c|}{} & 
	\multicolumn{4}{c|}{\cellcolor[HTML]{D9D9D9}... \textit{of which}} &  
	\multicolumn{3}{c||}{Bytes per data point} & 
	\multicolumn{1}{c|}{\cellcolor[HTML]{D9D9D9}} & 
	\multicolumn{1}{c|}{} &
	\multicolumn{1}{c|}{\cellcolor[HTML]{F2F2F2}} 
    
        \\ \cline{1-9}
	
	\rotatebox{90}{Num. of Points }			&
	\rotatebox{90}{Tetrahedra}							 		&
	\rotatebox{90}{\cellcolor[HTML]{D9D9D9}Peak Memory}\rotatebox{90}{\textcolor{gray}{{(MiB)}}}			&
   	\rotatebox{90}{Raw Data}										&
   	\rotatebox{90}{Construction}							 		&
	\rotatebox{90}{Simplification}							 		&
	\rotatebox{90}{Construction}									&
   	\rotatebox{90}{Simplification}									&
   	\rotatebox{90}{Total}											&
   	\rotatebox{90}{\cellcolor[HTML]{D9D9D9}Peak Memory}\rotatebox{90}{\textcolor{gray}{{(MiB)}}}			&
   	{\centering \raisebox{0.85\height}{\parbox{0.79cm}{Bytes\\ per\\ Data \\Point}}}	&
       	\rotatebox{90}{\cellcolor[HTML]{F2F2F2}Construction\textcolor{white}{,}}\rotatebox{90}{Memory Ratio}\rotatebox{90}{\textcolor{gray}{(\texttt{Viskores/TTK})}}

    \\	\cline{1-11}
10K  & 67K  & \cellcolor[HTML]{D9D9D9}21    & 0.23  & 16.10   & 4.7   & 1,610 & 490 & 2,100 & \cellcolor[HTML]{D9D9D9}40.29  & 4029 & \cellcolor[HTML]{F2F2F2}0.399 \\
30K  & 201K & \cellcolor[HTML]{D9D9D9}50.7  & 0.69  & 35.92   & 14.1  & 1,197 & 493 & 1,690 & \cellcolor[HTML]{D9D9D9}120.4  & 4013 & \cellcolor[HTML]{F2F2F2}0.298 \\
100K & 671K & \cellcolor[HTML]{D9D9D9}111.6 & 2.29  & 62.28   & 47.0  & 623   & 493 & 1,116 & \cellcolor[HTML]{D9D9D9}400.5  & 4005 & \cellcolor[HTML]{F2F2F2}0.155  \\
250K & 1.7M & \cellcolor[HTML]{D9D9D9}277.1 & 5.72  & 153.78  & 117.6 & 615   & 493 & 1,108 & \cellcolor[HTML]{D9D9D9}978	 & 3912 & \cellcolor[HTML]{F2F2F2}0.157  \\
1M   & 6.3M & \cellcolor[HTML]{D9D9D9}1,029 & 22.88 & 558.40  & 447.7 & 558   & 471 & 1,029 & \cellcolor[HTML]{D9D9D9}6,203  & 6203 & \cellcolor[HTML]{F2F2F2}0.090  \\
2.1M & 13M  & \cellcolor[HTML]{D9D9D9}2,160 & 49.44 & 1174.45 & 936.1 & 543   & 456 & 1,000 & \cellcolor[HTML]{D9D9D9}13,240 & 6127 & \cellcolor[HTML]{F2F2F2}0.089  \\
\cline{1-12}
\end{tabular}
    \caption{Comparison of \textit{memory usage} associated with contour tree analysis in \code{Viskores} (\code{VTK-m}) and \code{TTK} for an irregular mesh.}
    \label{tab:irregular-vtkm-ttk-memory}
\end{table*}

\begin{figure*}
    \centering
    \subfigure[$32^3$ grid (32K points)]{
        \includegraphics[width=0.33\linewidth]{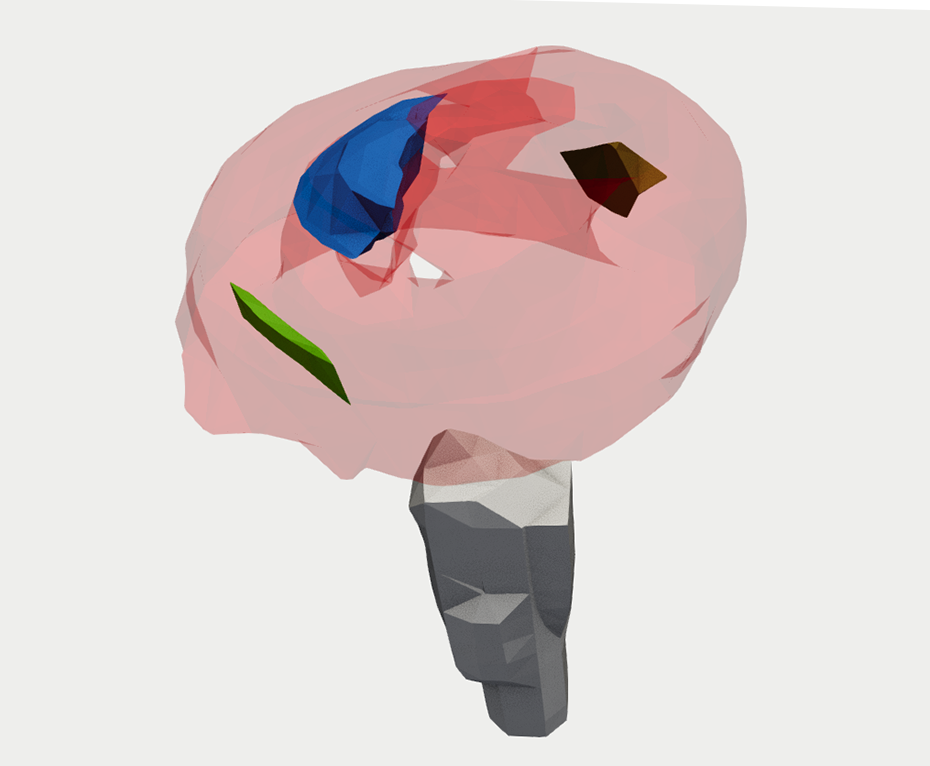}
        \label{fig:32k-grid}
    }\hspace*{-0.5em}
    \subfigure[$64^3$ grid (256K points)]{
        \includegraphics[width=0.33\linewidth]{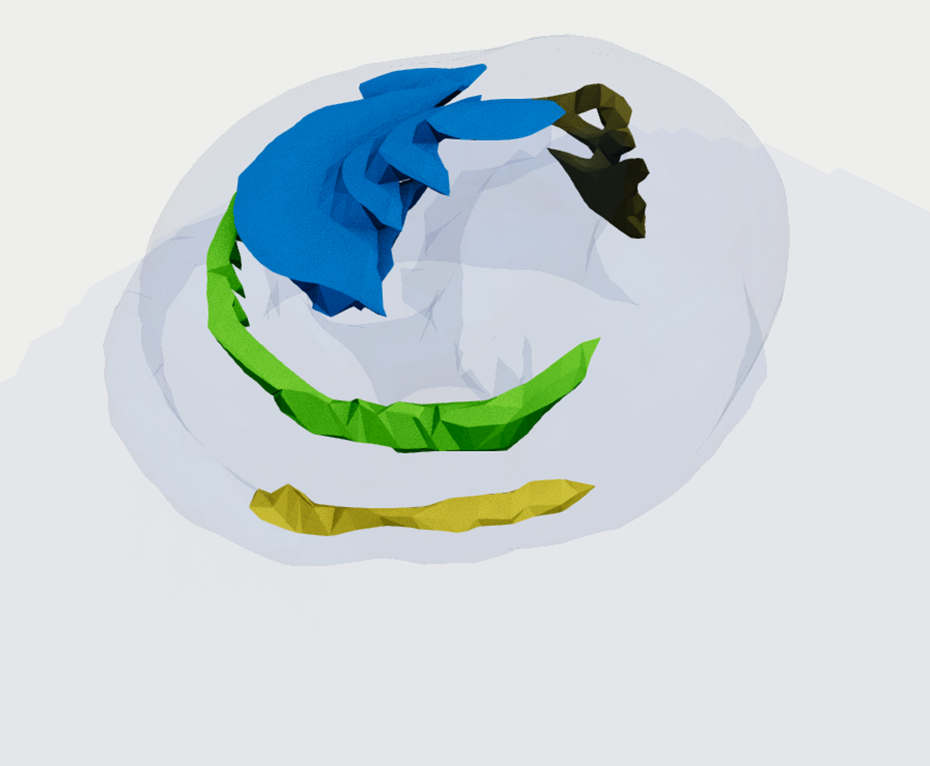}
        \label{fig:64k-grid}
    }\hspace*{-0.5em}
    \subfigure[$128^3$ grid (2M points)]{
        \includegraphics[width=0.33\linewidth]{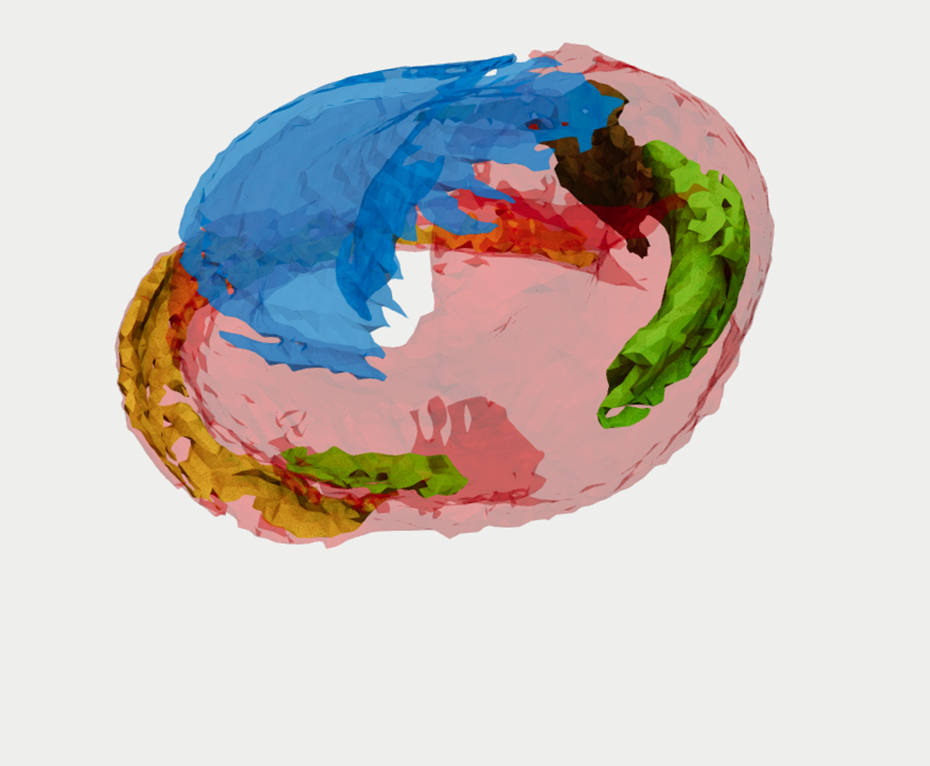}
        \label{fig:2m-grid}
    }

    \vspace{-1.1em}

    \subfigure[$256^3$ grid (16M points)]{
        \includegraphics[width=0.33\linewidth]{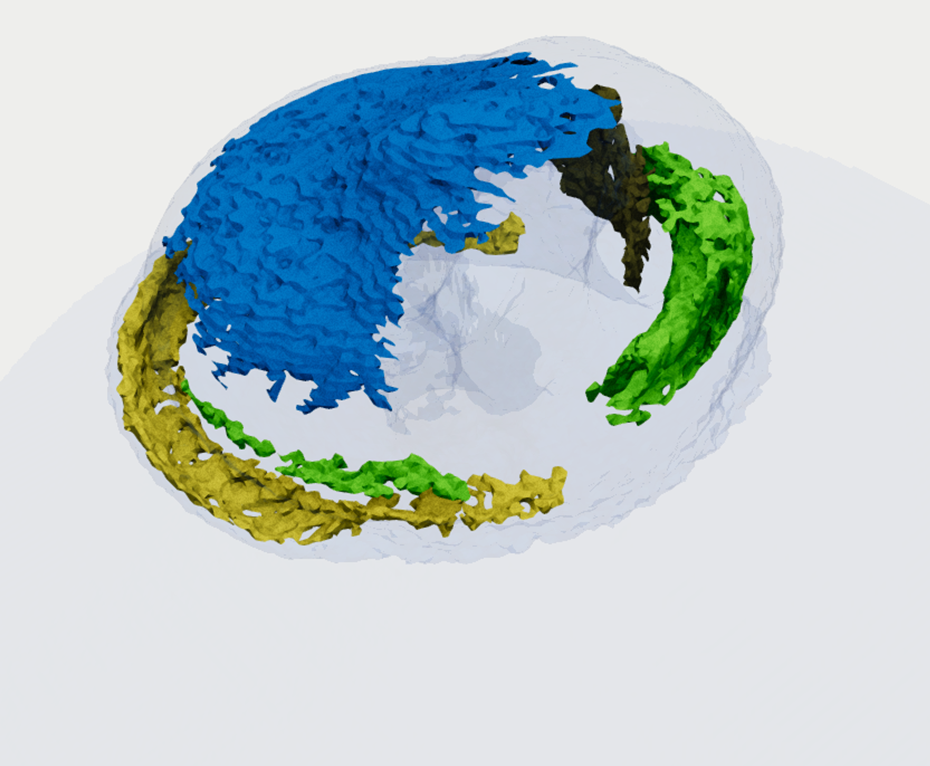}
        \label{fig:16m-grid}
    }\hspace*{-0.5em}
    \subfigure[$512^3$ grid (128M points)]{
        \includegraphics[width=0.33\linewidth]{GRID-0512_.png}
        \label{fig:128m-grid}
    }\hspace*{-0.5em}
    \subfigure[$1024^3$ grid (1B points)]{
        \includegraphics[width=0.33\linewidth]{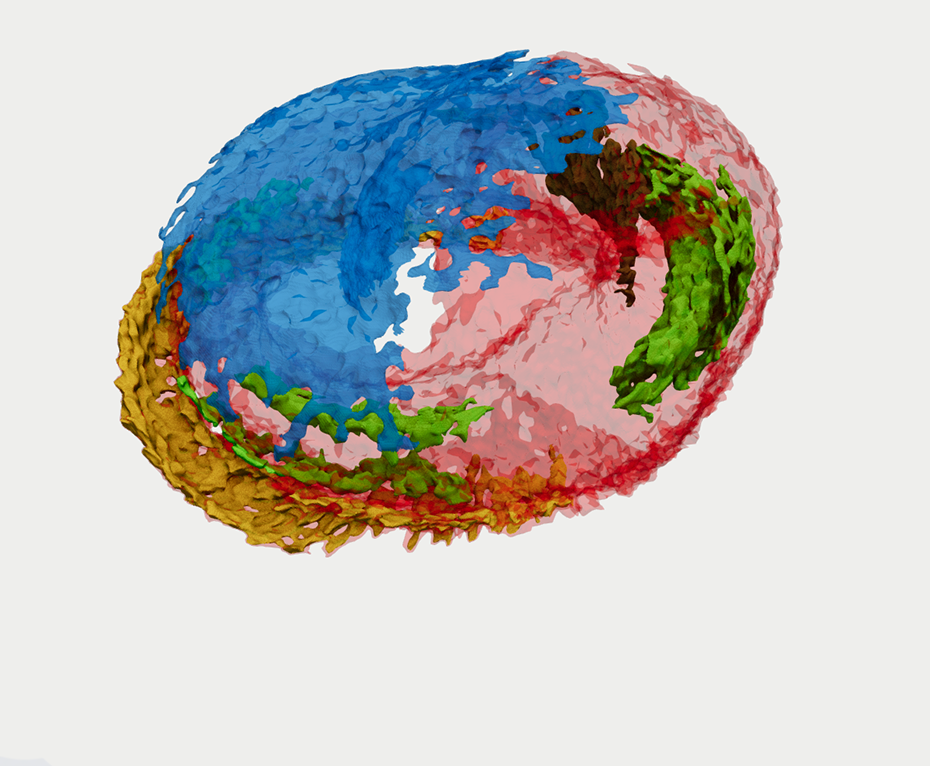}
        \label{fig:1b-grid}
    }

    \caption{Grid-based visualizations at different resolutions.}
    \label{fig:gridVisualization}
\end{figure*}

\paragraph{Methodology (Grid):}  The original $\sim$2 million parcel mesh-free simulation was first recomputed onto a uniform $1024^3$ mesh, following ~\cite{frey2022epic} Appendix H, and then coarse-grained to grid resolutions from $24^3$ to $512^3$. \xref{tab:grids-vtkm-ttk} shows the statistics collected, while \xref{fig:gridVisualization} shows  flexible isosurface visualizations for all resolutions. 

We recorded the data resolution, the number of input points (i.e. regular nodes), and the construction time for respective \code{TTK} and \code{Viskores} contour tree filters. For \code{Viskores}, we also show the time to count regular nodes for volume approximation, as well as the branch decomposition. We later use this to compare our new irregular mesh polynomial hypersweep, which computes \textit{exact} volume weights, and is more complex.
\code{TTK} outperforms \code{Viskores} for smaller resolutions, while larger resolutions take a significant amount of time ($> 16$ times slower than \code{Viskores}). 
This is not a surprise, since \code{TTK} targets workstations rather than many-core architectures and relies more on coarse-grain parallelism.  

We used \code{valgrind} to record peak memory usage.
At lower resolutions, \code{Viskores} uses half the memory of \code{TTK}, but converges for grids with more than $16M$ data points. 
We suspect this is because \code{TTK} switches to a fully implicit triangulation for meshes with $>16,777,216$ vertices. As the data scales, the size per data point ends up around $80-100B$ per sample point in both \code{Viskores} and \code{TTK}. 
At the highest resolution, \code{Viskores} is at the lower-end of this range ($86B$) compared to \code{TTK} ($101B$). For the smallest resolutions, the ratio is over $100$, but this is likely due to general overhead.

In \xref{fig:gridVisualization}, we see that the thermal column and cloud body are not reliably visible on grids until $128^3$ samples are available, but that a separate component at the top (the \textit{pileus}) is visible throughout. 
We also see that the vortex ring is hard to make out until higher resolutions, and tends to be visible as multiple components rather than the single toroidal shape we expect.

\begin{figure*}
    \centering
    \subfigure[PIC subsample (1K parcels)]{
        \includegraphics[width=0.33\linewidth]{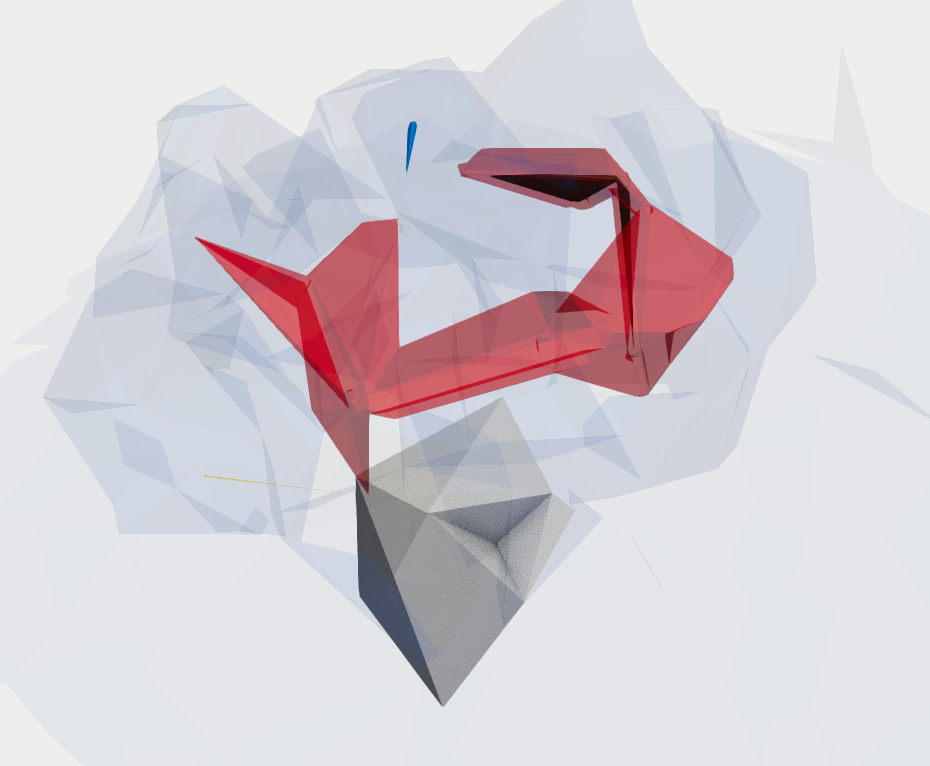}
        \label{fig:1K-parcels}
    }\hspace*{-0.5em}
    \subfigure[PIC subsample (10K parcels)]{
        \includegraphics[width=0.33\linewidth]{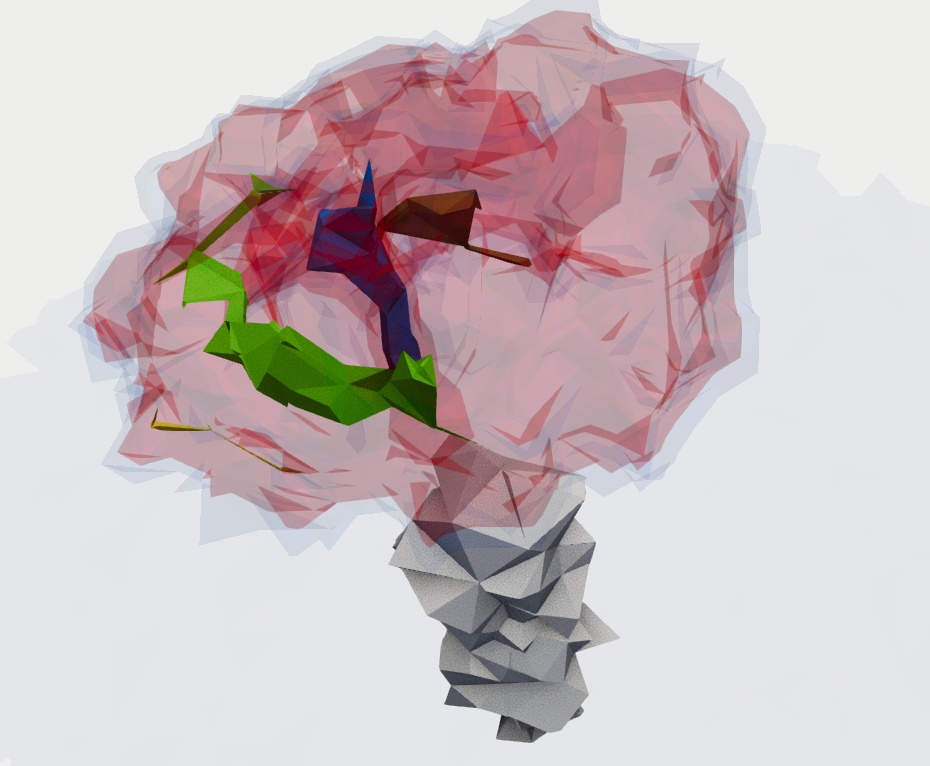}
        \label{fig:10K-parcels}
    }\hspace*{-0.5em}
    \subfigure[PIC subsample (100K parcels)]{
        \includegraphics[width=0.33\linewidth]{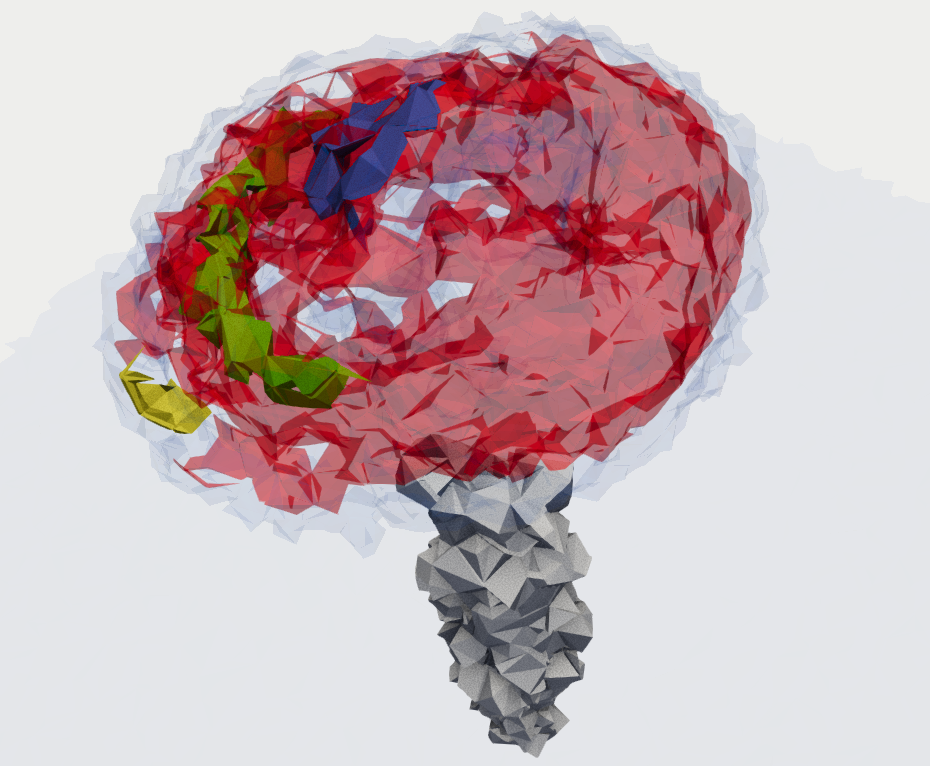}
        \label{fig:100K-parcels}
    }

    \vspace{-1.1em}

    \subfigure[PIC subsample (200K parcels)]{
        \includegraphics[width=0.33\linewidth]{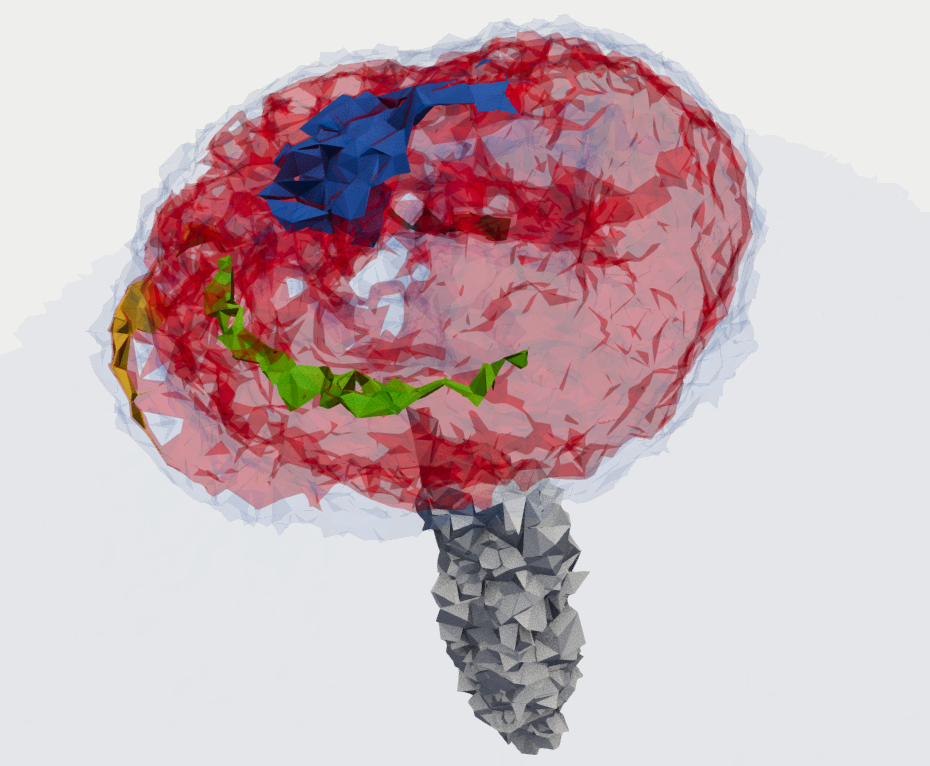}
        \label{fig:200K-parcels}
    }\hspace*{-0.5em}
    \subfigure[PIC subsample (1M parcels)]{
        \includegraphics[width=0.33\linewidth]{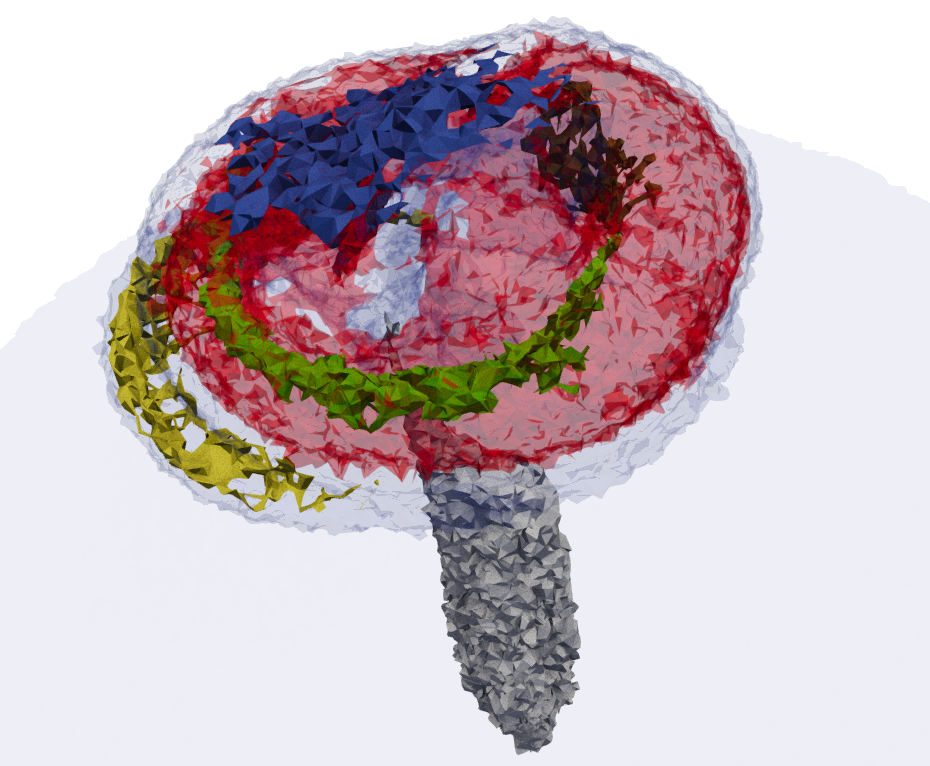}
        \label{fig:1M-parcels}
    }\hspace*{-0.5em}
    \subfigure[PIC - the original 2.1M parcels]{
        \includegraphics[width=0.33\linewidth]{PACTBDCV-2000K-semi.png}
        \label{fig:2M-parcels}
    }

    \caption{PIC-based visualizations based on random sub-selection of parcels.}
    \label{fig:PICVisualization}
\end{figure*}

\paragraph{Methodology (PIC):} We computed contour trees and exact volume weights from the EPIC simulation using the Delaunay pipeline (\xref{sec:implementation}), testing smaller data sizes with random down-samples of the parcels. We show visualizations in \xref{fig:PICVisualization} and compare the performance to \code{TTK} in tables \xref{tab:irregular-vtkm-ttk-timing}, \xref{tab:irregular-vtkm-ttk-memory} and plots \xref{fig:logplots}.

\xref{tab:irregular-vtkm-ttk-timing} shows the time costs of: Delaunay tetrahedralization, contour tree construction, the simplification using exact volume in \code{Viskores}, and computing the contour tree in \code{TTK} \new{(join/split trees and the merge)}. We do not report the simplification time cost in \code{TTK} because it does not have an exact volume computation. 

The Delaunay tetrahedralization is quasilinear~\cite{tetgen2016manual,hang2015tetgen} and dominates the time cost compared to the contour tree computation in both \code{Viskores} and \code{TTK}. However, even when we add all the total time steps: the tetrahedralization, the contour tree construction and (in \code{Viskores}) the exact volume simplification, the time cost is still smaller than for the highest resolution resampled grids ($1024^3$ in \code{Viskores} and both $512^3, 1024^3$ in \code{TTK}). Moreover, it took $\sim$23 \emph{minutes} to compute the $1024^3$ regridding of the original mesh-free simulation in the first place (compared to just 13 seconds for the Delaunay meshing on the original $\sim$2 million parcels).

We also break down the contour tree construction time. 
Both \code{Viskores} and \code{TTK} have two steps, building the edge list for each vertex, and running the contour tree constructor. In \code{Viskores},  the first step involves a prefix-sum style parallel worklet to extract the edge list from the delaunay mesh, while \code{TTK} uses a serial \code{OneSkeleton} extraction step. This is reflected in the timings, where \code{OneSkeleton} remains the dominant time cost for all resolutions. However, the serial extraction is undoubtedly parallel, so we primarily compare the main step, that of computing the contour tree itself.  In \code{Viskores}, we modified the existing \texttt{ContourTreeAugmented } worklet to handle arbitrary graph input (to support arbitrary vertex neighbourhood degrees), causing a significant slow-down compared to the original \code{Viskores} grid version and to \code{TTK}, as we can no longer rely on the optimisation that assumed $<32$ maximum neighbours per vertex (\xref{sec:implementation}). 

We note that the ratio of the \code{Viskores CTA} / \code{TTK FTMTree} filter timings is trending downwards with bigger datasets \xref{fig:logplots}(a) in much the same way as for grids. We predicted that around 30M parcels the \code{Viskores CTA} worklet would break even with the \code{FTMTree} filter. We ran a comparison for a larger, different, 30M parcel simulation, and measured that \code{Viskores/CTA} contour tree construction was twice as fast as \code{TTK/FTMTree} (\xref{fig:logplots}(a) - cross).

We then turn our attention to the time cost of exact volume computation in \code{Viskores}. Compared to grid-based volume approximation, this requires accessing the tetrahedral mesh during the hypersweep, and is in the region of $40-150$ times slower for the same number of regular nodes. 
This is because collecting the coefficients requires accessing the vertex coordinates and performing millions of double floating-point operations. 
While the computation of volume weights dominates the time cost, the branch decomposition occupies only a small fraction of the cost but is still about $3-4$ times slower than for the same number of input points on the grid. 
This is caused by the contour tree having an order of $10$ times more branches when computed from our EPIC Delaunay meshing.
Computation of volumetric weights is more expensive per vertex than for grids, but since we need to sum multiple sets of polynomial coefficients rather than just use the value $1$, this is not a surprise.

For memory usage,  \xref{tab:irregular-vtkm-ttk-memory} lists the number of sample parcels, the number of tetrahedra in the Delaunay mesh, the peak memory usage according to \code{valgrind} for both \code{Viskores} and \code{TTK}, and the overall cost per data point. For \code{Viskores}, we also report more fine-grained memory information, such as the memory for raw data and vertex positions, contour tree construction, and simplification.
Unsurprisingly, an explicit tetrahedral mesh requires more memory than an implicit gridded mesh with the same number of data points (\xref{tab:grids-vtkm-ttk}), but the memory footprint is around $1000B$ overall per data point, and the highest resolution (all $2.1M$ data points) occupies only $2GiB$ of memory, where the largest ($1024^3$) gridded data is $96GiB$.
In \code{TTK}, the memory footprint is around $4000B$ per data point for smaller resolutions, and $6000B$ for the largest, compared to $543B$ in \code{Viskores}. In \xref{fig:logplots}(b), we see that the \code{Viskores} / \code{TTK} ratio of memory footprint (for just the contour tree construction) is slightly trending downwards, consistently remaining below $1$.

\paragraph{Comparison:} Our original hypothesis was twofold: that contour tree analysis was more efficient and accurate without having to resample irregular meshes, and that flexible isosurface visualizations were potentially useful in studying cloud formation.

Looking at \xref{fig:ContoursVsTree}, the contour tree structures appear similar, except that we see more branches in the tree generated directly from the parcels because critical points are guaranteed to appear at the vertices \cite{carr2004topological}. The color-coded branches (which represent the largest features by volume) are labelled \textit{a-f} by order of volume and appear in similar regions in both datasets. The biggest features \textit{a, b, c} can be easily matched to well-known parts of a cloud, namely the vortex ring, the thermal column, and the pileus (the cap), in that order; \textit{d, e, f} are more difficult to pinpoint, especially since (\textit{d}, \textit{e}) appear as both topological peaks or pits, but look to be part of the vortex core - part of the cloud typically associated with strongest vorticity.

As we see in higher resolutions (\xref{fig:gridVisualization}, \xref{fig:PICVisualization}) we can reliably discern features of the cloud. In \xref{fig:gridVisualization} and \xref{fig:PICVisualization} we can see the quality improving as the resolution increases. 
However, one effect that we had not anticipated was the breaking-up of the vortex ring in gridded resampling, with the direct parcel-based version  producing more consistent results. 
We suspect that this is a side-effect of the interpolants used in the resampling, but were unable to test this.

In terms of efficiency, we can see that the direct computation with $2.1M$ parcels is time-competitive with gridded resamplings at around $384^3$,  and memory-competitive at around $256^3$ but that in the visualizations, the rising cloud stem and the vortex ring starts being visible even with as few as $10K$ parcels, and the $1M$ and $2M$ parcel versions appear to show not only the vortex ring itself but also the vortex core.  By the time that a regridded resolution starts showing the vortex core and ring reliably, however, the compute time is $2\times$ slower and the memory footprint $50\times$ greater. We also note that resampling $\sim$2 million parcels onto a $1024^3$ grid can take up to 30 \emph{minutes} even when using task-based parallelism (when using parcel resampling methods described in~\cite{frey2022epic} Appendix H), whereas the Delaunay tetrahedralization only takes about 13 seconds. Additionally, by doing the topological analysis directly on the parcels, the interpolation artifacts are avoided, such as the smoothing and blurring of the features; however, further studies on alternative neighbourhoods than just Delaunay are also of great interest. The downside of the coefficient-style exact volume computations is the summation of many very small floating point numbers, which is likely to cause precision errors for larger data, as we already noticed the dynamic range of the summation sequence reaching $1\mathrm{ e+}13$, prompting more future research for applying computer arithmetic methods for better numerical stability (especially in-parallel). 

We would therefore conclude that, where possible, both topological analysis and regular visualization of EPIC simulations is best done with the original parcels, after a meshing them with a Delaunay tetrahedralization rather than by resampling to a grid just to take advantage of existing tools.

\section{Summary and Future Work}
\label{sec:summary}

We have reported correctly developed polynomial coefficients for geometric properties in the well-known Contour Spectrum~\cite{BPS97}, and demonstrated how to compute them in PRAM parallel for data on an irregular mesh for contour tree analysis. This enables the computation of exact (i.e. not approximated) volumes for important regions of parcel-based scientific simulations. We have also reported on the implementation changes in the \code{Viskores} library necessary to add irregular mesh analysis to the existing methods.  We have shown that for a known data set, working directly with the underlying data representation is preferable to resampling just for the purpose of data visualization, and have shown that the efficiency gains in doing so can be several orders of magnitude. 

Several lines of inquiry now open up, in particular to return to Pascucci's computation of Betti numbers~\cite{Pas01}, but also to expand the hypersweep computation to a broader range of geometric measures~\cite{CSV10}, now that the basic mechanisms have been tested. 
Some more care will be needed to handle the numerical precision of this technique, which can become the limiting factor if not addressed.
Inevitably, though, this work will also need porting to the distributed layer, as the real value of topological analysis is to be found at the supercomputer scale rather than the single node which we used in our experiments.


\clearpage

\clearpage
\acknowledgments{
We thank the University of Leeds and UKRI for supporting the research (Grants: EP/W524372/1 DTP Studentship 2751073 and EP/T025409/1). This work was undertaken on the Aire HPC system at the University of Leeds, UK, with the simulation data produced on ARCHER2 UK National Supercomputing Service \cite{archer2}.
For technical assistance with flexible isosurfaces, we thank Petar Hristov. 
}

\bibliographystyle{abbrv-doi}

\bibliography{ms}

\begin{thebibliography}{10}

\bibitem{BPS97}
C.~L. Bajaj, V.~Pascucci, and D.~R. Schikore.
\newblock {The Contour Spectrum}.
\newblock In {\em Proceedings of Visualization 1997}, pp. 167--173, 1997. doi: {{%
10\hspace{.1pt}\discretionary{.}{%
}{.}\hspace{.4pt}1145\discretionary{/}{%
}{/}259081\hspace{.1pt}\discretionary{.}{%
}{.}\hspace{.4pt}259279}}


\bibitem{archer2}
G.~Beckett, J.~Beech-Brandt, K.~Leach, Z.~Payne, A.~Simpson, L.~Smith, A.~Turner, and A.~Whiting.
\newblock {ARCHER2 Service Description}, Dec. 2024. doi: {{%
10\hspace{.1pt}\discretionary{.}{%
}{.}\hspace{.4pt}5281\discretionary{/}{%
}{/}zenodo\hspace{.1pt}\discretionary{.}{%
}{.}\hspace{.4pt}14507040}}


\bibitem{blyth1980influence}
A.~Blyth, T.~Choularton, G.~Fullarton, J.~Latham, C.~Mill, M.~Smith, and I.~Stromberg.
\newblock {The Influence of Entrainment on the Evolution of Cloud Droplet Spectra .2. Field Experiments at Great-Dun-Fell}.
\newblock {\em Quarterly Journal of the Royal Meteorological Society}, 106(450):821--840, 1980. doi: {{%
10\hspace{.1pt}\discretionary{.}{%
}{.}\hspace{.4pt}1002\discretionary{/}{%
}{/}qj\hspace{.1pt}\discretionary{.}{%
}{.}\hspace{.4pt}49710645012}}


\bibitem{blyth1993entrainment}
A.~M. Blyth.
\newblock {Entrainment in Cumulus Clouds}.
\newblock {\em Journal of Applied Meteorology and Climatology}, 32(4):626--641, 1993. doi: {{%
10\hspace{.1pt}\discretionary{.}{%
}{.}\hspace{.4pt}1175\discretionary{/}{%
}{/}1520\discretionary{%
}{-}{-}0450\discretionary{%
}{(}{(}1993\discretionary{)}{%
}{)}032{\textless}0626\discretionary{:}{%
}{:}EICC{\textgreater}2\hspace{.1pt}\discretionary{.}{%
}{.}\hspace{.4pt}0\hspace{.1pt}\discretionary{.}{%
}{.}\hspace{.4pt}CO\discretionary{;}{%
}{;}2}}


\bibitem{boing2019comparison}
S.~J. B{\"o}ing, D.~G. Dritschel, D.~J. Parker, and A.~M. Blyth.
\newblock {Comparison of the Moist Parcel-in-Cell (MPIC) model with large-eddy simulation for an idealized cloud}.
\newblock {\em Quarterly Journal of the Royal Meteorological Society}, 145(722):1865--1881, 2019. doi: {{%
10\hspace{.1pt}\discretionary{.}{%
}{.}\hspace{.4pt}1002\discretionary{/}{%
}{/}qj\hspace{.1pt}\discretionary{.}{%
}{.}\hspace{.4pt}3532}}


\bibitem{carr2004topological}
H.~A. Carr.
\newblock {\em Topological Manipulation of Isosurfaces}.
\newblock PhD thesis, University of British Columbia, 2004. doi: {{%
10\hspace{.1pt}\discretionary{.}{%
}{.}\hspace{.4pt}14288\discretionary{/}{%
}{/}1\hspace{.1pt}\discretionary{.}{%
}{.}\hspace{.4pt}0051287}}


\bibitem{CRW22}
H.~A. Carr, O.~R{\"u}bel, and G.~H. Weber.
\newblock {Distributed Hierarchical Contour Trees}.
\newblock In {\em {IEEE Symposium on Large Data Analysis and Visualization (LDAV)}}, pp. 1--10, 2022. doi: {{%
10\hspace{.1pt}\discretionary{.}{%
}{.}\hspace{.4pt}1109\discretionary{/}{%
}{/}LDAV57265\hspace{.1pt}\discretionary{.}{%
}{.}\hspace{.4pt}2022\hspace{.1pt}\discretionary{.}{%
}{.}\hspace{.4pt}9966394}}


\bibitem{CWR22}
H.~A. Carr, O.~R{\"u}bel, and G.~H. Weber.
\newblock {Distributed Hierarchical Contour Trees}.
\newblock In {\em 2022 IEEE 12th Symposium on Large Data Analysis and Visualization (LDAV)}, pp. 1--10. IEEE, 2022. doi: {{%
10\hspace{.1pt}\discretionary{.}{%
}{.}\hspace{.4pt}1109\discretionary{/}{%
}{/}LDAV57265\hspace{.1pt}\discretionary{.}{%
}{.}\hspace{.4pt}2022\hspace{.1pt}\discretionary{.}{%
}{.}\hspace{.4pt}9966394}}


\bibitem{CS09}
H.~A. Carr and J.~Snoeyink.
\newblock {Representing Interpolant Topology for Contour Tree Computation}.
\newblock In H.-C. Hege, K.~Polthier, and G.~Scheuermann, eds., {\em Topology-Based Methods in Visualization II}, Mathematics and Visualization, pp. 59--74. Springer, 2009. doi: {{%
10\hspace{.1pt}\discretionary{.}{%
}{.}\hspace{.4pt}1007\discretionary{/}{%
}{/}978\discretionary{%
}{-}{-}3\discretionary{%
}{-}{-}540\discretionary{%
}{-}{-}88606\discretionary{%
}{-}{-}8\_5}}


\bibitem{carr2004simplifying}
H.~A. Carr, J.~Snoeyink, and M.~van~de Panne.
\newblock {Simplifying Flexible Isosurfaces Using Local Geometric Measures}.
\newblock In {\em IEEE Visualization 2004}, pp. 497--504, 2004. doi: {{%
10\hspace{.1pt}\discretionary{.}{%
}{.}\hspace{.4pt}1109\discretionary{/}{%
}{/}VISUAL\hspace{.1pt}\discretionary{.}{%
}{.}\hspace{.4pt}2004\hspace{.1pt}\discretionary{.}{%
}{.}\hspace{.4pt}96}}


\bibitem{CSV10}
H.~A. Carr, J.~Snoeyink, and M.~van~de Panne.
\newblock {Flexible Isosurfaces: Simplifying and Displaying Scalar Topology Using the Contour Tree}.
\newblock {\em {Computational Geometry: Theory and Applications (CGTA)}}, 43(1):42--58, 2010. doi: {{%
10\hspace{.1pt}\discretionary{.}{%
}{.}\hspace{.4pt}1016\discretionary{/}{%
}{/}j\hspace{.1pt}\discretionary{.}{%
}{.}\hspace{.4pt}comgeo\hspace{.1pt}\discretionary{.}{%
}{.}\hspace{.4pt}2006\hspace{.1pt}\discretionary{.}{%
}{.}\hspace{.4pt}05\hspace{.1pt}\discretionary{.}{%
}{.}\hspace{.4pt}009}}


\bibitem{CWS16}
H.~A. Carr, G.~H. Weber, C.~Sewell, and J.~Ahrens.
\newblock {Parallel Peak Pruning for Scalable SMP Contour Tree Computation}.
\newblock In {\em {IEEE Symposium on Large Data Analysis and Visualization (LDAV)}}, pp. 75--84, 2016. doi: {{%
10\hspace{.1pt}\discretionary{.}{%
}{.}\hspace{.4pt}1109\discretionary{/}{%
}{/}LDAV\hspace{.1pt}\discretionary{.}{%
}{.}\hspace{.4pt}2016\hspace{.1pt}\discretionary{.}{%
}{.}\hspace{.4pt}7874312}}


\bibitem{correa2011towards}
C.~Correa and P.~Lindstrom.
\newblock Towards robust topology of sparsely sampled data.
\newblock {\em IEEE Transactions on Visualization and Computer Graphics}, 17(12):1852--1861, 2011. doi: {{%
10\hspace{.1pt}\discretionary{.}{%
}{.}\hspace{.4pt}1109\discretionary{/}{%
}{/}TVCG\hspace{.1pt}\discretionary{.}{%
}{.}\hspace{.4pt}2011\hspace{.1pt}\discretionary{.}{%
}{.}\hspace{.4pt}245}}


\bibitem{dabiri2004fluid}
J.~O. Dabiri and M.~Gharib.
\newblock Fluid entrainment by isolated vortex rings.
\newblock {\em Journal of fluid mechanics}, 511:311--331, 2004. doi: {{%
10\hspace{.1pt}\discretionary{.}{%
}{.}\hspace{.4pt}1017\discretionary{/}{%
}{/}S0022112004009784}}


\bibitem{deardorff1970three}
J.~W. Deardorff.
\newblock A three-dimensional numerical investigation of the idealized planetary boundary layer.
\newblock {\em Geophysical and Astrophysical Fluid Dynamics}, 1(3-4):377--410, 1970. doi: {{%
10\hspace{.1pt}\discretionary{.}{%
}{.}\hspace{.4pt}1080\discretionary{/}{%
}{/}03091927009365780}}


\bibitem{Drebin1988}
R.~A. Drebin, L.~Carpenter, and P.~Hanrahan.
\newblock Volume rendering.
\newblock In {\em Proceedings of the 15th Annual Conference on Computer Graphics and Interactive Techniques}, {SIGGRAPH} '88, pp. 65--74. Association for Computing Machinery, New York, NY, USA, 1988. doi: {{%
10\hspace{.1pt}\discretionary{.}{%
}{.}\hspace{.4pt}1145\discretionary{/}{%
}{/}54852\hspace{.1pt}\discretionary{.}{%
}{.}\hspace{.4pt}378484}}


\bibitem{DCM12}
B.~Duffy, H.~A. Carr, and T.~M{\"o}ller.
\newblock {Integrating Isosurface Statistics and Histograms}.
\newblock {\em {IEEE Transactions on Visualization and Computer Graphics (TVCG)}}, 19(2):263--277, 2012. doi: {{%
10\hspace{.1pt}\discretionary{.}{%
}{.}\hspace{.4pt}1109\discretionary{/}{%
}{/}TVCG\hspace{.1pt}\discretionary{.}{%
}{.}\hspace{.4pt}2012\hspace{.1pt}\discretionary{.}{%
}{.}\hspace{.4pt}118}}


\bibitem{federer_gmt_69}
H.~Federer.
\newblock {\em {Geometric Measure Theory}}.
\newblock Springer-Verlag, 1965. doi: {{%
10\hspace{.1pt}\discretionary{.}{%
}{.}\hspace{.4pt}1007\discretionary{/}{%
}{/}978\discretionary{%
}{-}{-}3\discretionary{%
}{-}{-}642\discretionary{%
}{-}{-}62010\discretionary{%
}{-}{-}2}}


\bibitem{frey2022epic}
M.~Frey, D.~Dritschel, and S.~B{\"o}ing.
\newblock {EPIC: the Elliptical Parcel-In-Cell method}.
\newblock {\em Journal of Computational Physics: X}, 14:100109, 2022. doi: {{%
10\hspace{.1pt}\discretionary{.}{%
}{.}\hspace{.4pt}1016\discretionary{/}{%
}{/}j\hspace{.1pt}\discretionary{.}{%
}{.}\hspace{.4pt}jcpx\hspace{.1pt}\discretionary{.}{%
}{.}\hspace{.4pt}2022\hspace{.1pt}\discretionary{.}{%
}{.}\hspace{.4pt}100109}}


\bibitem{frey20233d}
M.~Frey, D.~Dritschel, and S.~B{\"o}ing.
\newblock {The 3D Elliptical Parcel-In-Cell (EPIC) method}.
\newblock {\em Journal of Computational Physics: X}, 17:100136, 2023. doi: {{%
10\hspace{.1pt}\discretionary{.}{%
}{.}\hspace{.4pt}1016\discretionary{/}{%
}{/}j\hspace{.1pt}\discretionary{.}{%
}{.}\hspace{.4pt}jcpx\hspace{.1pt}\discretionary{.}{%
}{.}\hspace{.4pt}2023\hspace{.1pt}\discretionary{.}{%
}{.}\hspace{.4pt}100136}}


\bibitem{GFJ16}
C.~Gueunet, P.~Fortin, and J.~Jomier.
\newblock {Contour forests: Fast multi-threaded augmented contour trees}.
\newblock In {\em 6th IEEE Symposium on Large Data Analysis and Visualization (LDAV)}, pp. 85--92, Oct 2016. doi: {{%
10\hspace{.1pt}\discretionary{.}{%
}{.}\hspace{.4pt}1109\discretionary{/}{%
}{/}LDAV\hspace{.1pt}\discretionary{.}{%
}{.}\hspace{.4pt}2016\hspace{.1pt}\discretionary{.}{%
}{.}\hspace{.4pt}7874333}}


\bibitem{HWC20}
P.~Hristov, G.~H. Weber, H.~A. Carr, O.~R{\"u}bel, and J.~Ahrens.
\newblock {Data Parallel Hypersweeps for in Situ Topological Analysis}.
\newblock In {\em {IEEE Symposium on Large Data Analysis and Visualization (LDAV)}}, pp. 12--21, 2020. doi: {{%
10\hspace{.1pt}\discretionary{.}{%
}{.}\hspace{.4pt}1109\discretionary{/}{%
}{/}LDAV51489\hspace{.1pt}\discretionary{.}{%
}{.}\hspace{.4pt}2020\hspace{.1pt}\discretionary{.}{%
}{.}\hspace{.4pt}00008}}


\bibitem{levoy1988display}
M.~Levoy.
\newblock Display of surfaces from volume data.
\newblock {\em {IEEE Computer Graphics and Applications}}, 8(3):29--37, 1988. doi: {{%
10\hspace{.1pt}\discretionary{.}{%
}{.}\hspace{.4pt}1109\discretionary{/}{%
}{/}38\hspace{.1pt}\discretionary{.}{%
}{.}\hspace{.4pt}511}}


\bibitem{lorensen1987marching}
W.~E. Lorensen and H.~E. Cline.
\newblock {Marching cubes: A high resolution 3D surface construction algorithm}.
\newblock {\em ACM SIGGRAPH Computer Graphics}, 21(4):163--169, 1987. doi: {{%
10\hspace{.1pt}\discretionary{.}{%
}{.}\hspace{.4pt}1145\discretionary{/}{%
}{/}37402\hspace{.1pt}\discretionary{.}{%
}{.}\hspace{.4pt}37422}}


\bibitem{mellado2017cloud}
J.~P. Mellado.
\newblock {Cloud-Top Entrainment in Stratocumulus Clouds}.
\newblock {\em Annual Review of Fluid Mechanics}, 49:145--169, 2017. doi: {{%
10\hspace{.1pt}\discretionary{.}{%
}{.}\hspace{.4pt}1146\discretionary{/}{%
}{/}annurev\discretionary{%
}{-}{-}fluid\discretionary{%
}{-}{-}010816\discretionary{%
}{-}{-}060231}}


\bibitem{MR85}
G.~L. Miller and J.~H. Reif.
\newblock {Parallel Tree Contraction and Its Application}.
\newblock In {\em IEEE Symposium on Foundations of Computer Science (FOCS)}, vol.~26, pp. 478--489, 1985. doi: {{%
10\hspace{.1pt}\discretionary{.}{%
}{.}\hspace{.4pt}1109\discretionary{/}{%
}{/}SFCS\hspace{.1pt}\discretionary{.}{%
}{.}\hspace{.4pt}1985\hspace{.1pt}\discretionary{.}{%
}{.}\hspace{.4pt}43}}


\bibitem{MSU16}
K.~Moreland, C.~Sewell, W.~Usher, L.-T. Lo, J.~Meredith, D.~Pugmire, J.~Kress, H.~Schroots, K.~L. Ma, H.~Childs, M.~Larsen, C.~M. Chen, R.~Maynard, and B.~Geveci.
\newblock {VTK-m: Accelerating the Visualization Toolkit for Massively Threaded Architectures}.
\newblock {\em {IEEE Computer Graphics and Applications}}, 36(3):48--58, May 2016. doi: {{%
10\hspace{.1pt}\discretionary{.}{%
}{.}\hspace{.4pt}1109\discretionary{/}{%
}{/}MCG\hspace{.1pt}\discretionary{.}{%
}{.}\hspace{.4pt}2016\hspace{.1pt}\discretionary{.}{%
}{.}\hspace{.4pt}48}}


\bibitem{Pas01}
V.~Pascucci.
\newblock {On the Topology of the Level Sets of a Scalar Field}.
\newblock In {\em Abstracts of the 13th Canadian Conference on Computational Geometry}, pp. 141--144, 2001.

\bibitem{PC03}
V.~Pascucci and K.~Cole-McLaughlin.
\newblock {Parallel Computation of the Topology of Level Sets}.
\newblock {\em Algorithmica}, 38(2):249--268, 2003. doi: {{%
10\hspace{.1pt}\discretionary{.}{%
}{.}\hspace{.4pt}1007\discretionary{/}{%
}{/}s00453\discretionary{%
}{-}{-}003\discretionary{%
}{-}{-}1052\discretionary{%
}{-}{-}3}}


\bibitem{PCS04}
V.~Pascucci, K.~Cole-McLaughlin, and G.~Scorzell.
\newblock {Multi-Resolution computation and presentation of Contour Trees}.
\newblock In {\em Proceedings of the IASTED conference on Visualization, Imaging and Image Processing (VIIP 2004)}, pp. 452--290, 2004.

\bibitem{Ree46}
G.~Reeb.
\newblock {Sur les points singuliers d'une forme de Pfaff compl{\`e}tement int{\'e}grable ou d'une fonction num{\'e}rique.}
\newblock {\em Comptes Rendus de l'Acad{\`e}mie des Sciences de Paris}, 222:847--849, 1946. doi: {{%
10\hspace{.1pt}\discretionary{.}{%
}{.}\hspace{.4pt}24033\discretionary{/}{%
}{/}asens\hspace{.1pt}\discretionary{.}{%
}{.}\hspace{.4pt}776}}


\bibitem{SSD08}
C.~E. Scheidegger, J.~M. Schreiner, B.~Duffy, H.~A. Carr, and C.~T. Silva.
\newblock {Revisiting Histograms and Isosurface Statistics}.
\newblock {\em {IEEE Transactions on Visualization and Computer Graphics}}, 14(6):1659--1666, 2008. doi: {{%
10\hspace{.1pt}\discretionary{.}{%
}{.}\hspace{.4pt}1109\discretionary{/}{%
}{/}TVCG\hspace{.1pt}\discretionary{.}{%
}{.}\hspace{.4pt}2008\hspace{.1pt}\discretionary{.}{%
}{.}\hspace{.4pt}160}}


\bibitem{shannon1949communication}
C.~E. Shannon.
\newblock {Communication in the Presence of Noise}.
\newblock {\em Proceedings of the IRE}, 37(1):10--21, 1949. doi: {{%
10\hspace{.1pt}\discretionary{.}{%
}{.}\hspace{.4pt}1109\discretionary{/}{%
}{/}JRPROC\hspace{.1pt}\discretionary{.}{%
}{.}\hspace{.4pt}1949\hspace{.1pt}\discretionary{.}{%
}{.}\hspace{.4pt}232969}}


\bibitem{hang2015tetgen}
H.~Si.
\newblock {TetGen A Delaunay-Based Quality Tetrahedral Mesh Generator}.
\newblock {\em ACM Trans. Math. Softw}, 41(2):11, 2015. doi: {{%
10\hspace{.1pt}\discretionary{.}{%
}{.}\hspace{.4pt}13140\discretionary{/}{%
}{/}RG\hspace{.1pt}\discretionary{.}{%
}{.}\hspace{.4pt}2\hspace{.1pt}\discretionary{.}{%
}{.}\hspace{.4pt}2\hspace{.1pt}\discretionary{.}{%
}{.}\hspace{.4pt}13915\hspace{.1pt}\discretionary{.}{%
}{.}\hspace{.4pt}85284\discretionary{/}{%
}{/}2}}


\bibitem{tetgen2016manual}
H.~Si.
\newblock {\em {TetGen A Quality Tetrahedral Mesh Generator and Three-Dimensional Delaunay Triangulator}}, 2020.

\bibitem{TFL17}
J.~Tierny, G.~Favelier, J.~A. Levine, C.~Gueunet, and M.~Michaux.
\newblock {The Topology ToolKit}.
\newblock {\em IEEE Transactions on Visualization and Computer Graphics}, 24(1):832--842, 2017. doi: {{%
10\hspace{.1pt}\discretionary{.}{%
}{.}\hspace{.4pt}48550\discretionary{/}{%
}{/}arXiv\hspace{.1pt}\discretionary{.}{%
}{.}\hspace{.4pt}1805\hspace{.1pt}\discretionary{.}{%
}{.}\hspace{.4pt}09110}}


\bibitem{zhou2018efficient}
B.~Zhou, Y.-J. Chiang, and C.~Wang.
\newblock {Efficient Local Statistical Analysis via Point-Wise Histograms in Tetrahedral Meshes and Curvilinear Grids}.
\newblock {\em IEEE Transactions on Visualization and Computer Graphics}, 25(2):1392--1406, 2018. doi: {{%
10\hspace{.1pt}\discretionary{.}{%
}{.}\hspace{.4pt}1109\discretionary{/}{%
}{/}TVCG\hspace{.1pt}\discretionary{.}{%
}{.}\hspace{.4pt}2018\hspace{.1pt}\discretionary{.}{%
}{.}\hspace{.4pt}2796555}}


\end{thebibliography}
\end{document}